# Nonreciprocal superconducting critical currents with normal state field trainability in kagome superconductor $CsV_3Sb_5$

Jun Ge[1#], Xiaoqi Liu[1#], Pinyuan Wang[1], Haowen Pang[1], Qiangwei Yin[3], Hechang Lei[3], Ziqiang Wang[4] & Jian Wang[1,2,5,6*]

[1]*International Center for Quantum Materials, School of Physics, Peking University, Beijing 100871, China*
[2]*Beijing Key Laboratory of Quantum Devices, Peking University, Beijing 100871, China*
[3]*Beijing Key Laboratory of Optoelectronic Functional Materials & Micro-Nano Devices, Department of Physics, Renmin University of China, Beijing 100872, China*
[4]*Department of Physics, Boston College, Chestnut Hill, MA 0246, USA*
[5]*Collaborative Innovation Center of Quantum Matter, Beijing 100871, China*
[6]*Hefei National Laboratory, Hefei 230088, China*

[#]These authors contribute equally.
*Correspondence to: jianwangphysics@pku.edu.cn (J.W.)

**Determining time-reversal symmetry (TRS) and chirality in the superconducting state and its relation to normal-state symmetry and topology are important issues in condensed matter physics. Here, we report nonreciprocal superconducting critical currents ($I_c$) at zero magnetic field in kagome superconductor $CsV_3Sb_5$ nanodevices: $I_c$ differs for opposite directions, indicating spontaneous TRS and inversion symmetry breakings. The polarity of $I_c$ asymmetry changes randomly in repeated thermal cycling to 300 K, consistent with spontaneous TRS breaking. Crucially, on applying a perpendicular magnetic field above the charge density wave (CDW) transition temperature and then removing it to zero above the superconducting onset temperature ($T_c$), the polarity of $I_c$ asymmetry follows the field direction, ascertaining that the CDW state has a macroscopic and trainable TRS-breaking directionality. The symmetry breaking continues into the superconducting state and generates the nonreciprocal critical currents. These results provide evidence for the loop-current CDW normal state with TRS breaking in $CsV_3Sb_5$.**

## Introduction

The layered-structure kagome superconductors $AV_3Sb_5$ (A= K, Rb, Cs) have attracted enormous research interests. Utilizing the combined effects of geometric frustration of the kagome lattice, electron correlations, and electron-lattice interaction on the kagome network, a wide variety of correlated and topological electronic states have been proposed for the compounds[1–3]. The compounds have a high temperature charge density wave transition at $T_{CDW}$ varies from 84 K to 104 K[4–7], below which the translational symmetry changes and a superconducting transition at 0.9 K-2.5 K arises[5–7]. Among the $AV_3Sb_5$ (A= K, Rb, Cs) compounds, $CsV_3Sb_5$ possesses the highest critical temperature of superconductivity ($T_c$~2.5 K)[4,5] and stands out as the most promising platform for studying the interplay between superconductivity, unconventional charge orders in the normal state, and band topology. Bulk $CsV_3Sb_5$ crystallizes into the P6/mmm space group. As shown in Fig. 1a, vanadium (V) cations form a two-dimensional (2D) kagome network[4,5]. Adjacent V kagome sheets are separated by layers of Cs ions. There are proposals, based on approximate calculations, that $CsV_3Sb_5$ below $T_{CDW}$ is a doped orbital Chern insulator near van Hove filling[8–10] due to the TRS breaking, i.e. persistent electrical loop-current in a complex 3Q CDW order generated by Coulomb interactions.

Various intriguing quantum phenomena have been observed in $CsV_3Sb_5$, including the charge-4*e* and charge-6*e* superconductivity[11], the pair density wave (PDW)[12,13], the stripe CDW[14], the nematic order[15–19], the chiral CDW order[20–23], giant anomalous Hall effect[24], and indications of TRS breaking in the CDW state[16,18,21–23,25–27] and the superconducting state[13,26,28] in experiments such as muon spin-rotation[21,25,26], scanning tunneling microscopy[13,22,23] (STM), optical[16,18,27] and transport measurements[28,29]. However, whether TRS is genuinely broken in both the high temperature CDW order and the superconducting order remains controversial due to the contradictory results[30–33], and the evidence connecting the CDW order to the superconducting state is still lacking.

In this work, we provide macroscopic evidence for TRS breaking by observing unambiguous nonreciprocal critical currents in zero-field, i.e. the spontaneous superconducting diode effect (SDE), which settles some debated issues and supports TRS breaking in $CsV_3Sb_5$. The polarity reversal of the SDE under zero-field thermal cycles indicates the existence of the superconducting domains with different chirality in the samples and further confirms the chiral superconductivity in $CsV_3Sb_5$[8,13,28,34,35]. Strikingly, the polarity of the SDE in the superconducting state can be changed by reversing the magnetic field in the normal state, revealing the intertwined relation between the normal state with CDW orders and the superconducting state. Our observations of the nonreciprocal critical currents and the field-training-tuned polarity in $CsV_3Sb_5$ indicate that the TRS breaking starts in the CDW state and continues into the superconducting state in $CsV_3Sb_5$.

## Results

**Nonreciprocal critical currents at zero-field in $CsV_3Sb_5$ thin flake devices**

We fabricate seven $CsV_3Sb_5$ thin flake devices (f1-f7), which exhibit low contact resistance (Supplementary Table 1), ensuring reliable transport characterization. We employ rigorous protocols to eliminate residual magnetic fields before all zero-field measurements. For devices f1 and f2, measured in a Re-liquefier based 16 T PPMS (Quantum Design), the magnet remanence is carefully eliminated by oscillating the magnetic field to zero at room temperature prior to cooling. For device f3, measured in a cryogen-free PPMS Dynacool (14 T), the superconducting magnet is warmed to room temperature to fully release any trapped magnetic flux. The zero-field environment is further verified using a high-precision Hall sensor (HGCT 3020, Lakeshore). For devices f4-f7, in addition to applying a compensating magnetic field to achieve a zero-field condition (typically below 0.01 Oe), as calibrated by a high-precision Hall sensor, we further perform *in-situ* device flipping under zero field to exclude the possible influence of any residual magnetic field (see Methods). Supplementary Figs. 1a-g display the resistance ($R$)-temperature ($T$) curves from 300 K to 2 K for the flake devices. The kink features in the $R$-$T$ curves at around 82 K, 74 K, 78 K, 70 K, 80 K, 76 K and 75 K are detected in f1-f7, respectively, indicating the CDW transitions in these thin flake devices. Figure 1b displays the $R$-$T$ curve from 4 K to 1.2 K in zero magnetic field of $CsV_3Sb_5$ flake device f2 with thickness of ~ 8.8 nm, with the atomic force microscopy (AFM) results in Supplementary Fig. 2b. The superconducting transition with the onset temperature $T_c^{onset}$ ~ 3.50 K and the zero-resistance temperature $T_c^{zero}$ ~3.18 K is detected and illustrated in Fig. 1b.

We then investigate the voltage ($V$)-current ($I$) curves of the $CsV_3Sb_5$ flake devices at zero magnetic field. The current is slowly ramped from negative to positive (defined as positive sweep) and from positive to negative (defined as negative sweep). The typical $V$-$I$ curves in flake device f2 at 0.9 K are shown in Fig. 1c. Note that the forward (current sweeping from zero to positive/negative) and return traces (current sweeping from positive/negative to zero) overlap with each other (Fig. 1c), indicating that Joule heating is negligible. Focused on the 0-P branch (current ramping from zero to positive) of the positive sweep, the zero-resistance superconducting state is observed below $I_{c+}$ ~ 139 μA. With increasing current above $I_{c+}$~ 139 μA, the voltage sharply jumps from zero to a finite value and the zero-resistance superconducting state disappears. On the 0-N branch (current ramping from zero to negative), the voltage sharply jumps to the resistive state at a critical current $I_{c-}$~ -137 μA. Obviously, nonreciprocal critical currents $|I_{c-}| \neq |I_{c+}|$ are detected through direct measurements of $V$-$I$ curves under the zero-field condition in the $CsV_3Sb_5$ flake device. For a better comparison between the critical currents in opposite directions, in Fig. 1d, we plot the absolute $V$-$I$ values in 0-P and 0-N branches derived from Fig. 1c. A clear imbalance between $I_{c+}$ and $|I_{c-}|$ can be directly observed from the $V$-$I$ curves, suggesting an intrinsic nonreciprocity of critical currents in the $CsV_3Sb_5$ flake device.

The observed imbalance of the critical currents in opposite directions indicates the emergence of the SDE. The SDE, characterized by the different critical currents along opposite directions[36–44], is commonly believed to originate from the TRS and inversion symmetry breakings in superconductors[45–51]. Up to now, the SDE has been observed in superlattices[39,52], heterostructures[41], Josephson junctions[41–43,53–56] and thin films[44,57,58]. In these systems, the inversion symmetry breaking typically arises from the geometry asymmetry[39,44,54–57,59,60] and the TRS breaking is usually realized by applying external magnetic fields[39,44,57,61] or introducing magnetic layers[38,52,53,62]. Previous works also show that the SDE can be achieved without magnetic field or magnetic layer[28,40,41,63]. The difference of $I_{c+}$ and $|I_{c-}|$ shown in Fig. 1d implies that if the applied excitation current lies between $I_{c+}$ and $|I_{c-}|$, the $CsV_3Sb_5$ flake device will be in the zero-resistance superconducting state for current flowing along the positive direction but in a resistive state for current along the opposite direction. Switching the positive and negative directions of the current with amplitude between $I_{c+}$ and $|I_{c-}|$, e.g. by applying a square-wave excitation, the resistance will switch between zero and a finite value, giving rise to half-wave the rectification effect. Considering the $I_{c+}$ and $|I_{c-}|$ being 139 and 137 μA, respectively, we apply a square-wave excitation (Fig. 1e, upper panel) with an amplitude of 138 μA at a frequency of 0.0079 Hz to confirm the existence of rectification effect under zero magnetic field. As shown in the lower panel of Fig. 1e, half-wave rectification is beautifully observed as the voltage switches between zero in superconducting state and a finite value in the resistive state. The half-wave rectification at other temperatures is shown in Supplementary Fig. 3. The zero-field SDE is also observed in other $CsV_3Sb_5$ flake devices (Supplementary Figs. 4-6). The maximum diode efficiency $\eta$ ($\eta = \frac{I_{c+} - |I_{c-}|}{I_{c+} + |I_{c-}|} \times 100\%$) reaches 17% in f1 and 22% in f3.

Figure 1f shows the temperature dependence of the $I_{c+}$, $|I_{c-}|$ (the upper panel), $\Delta I_c$ ($I_{c+}$-$|I_{c-}|$, the middle panel), and the diode efficiency $\eta$ (the lower panel) in $CsV_3Sb_5$ flake device f2. As the temperature increases from 0.7 K to 1.95 K, $I_{c+}$ remains larger than $|I_{c-}|$. When the temperature is above 2 K, $I_{c+}$ becomes smaller than $|I_{c-}|$, i.e. the polarity of the SDE reverses. The polarity reversal with increasing temperatures is also observed in other $CsV_3Sb_5$ flake devices (Supplementary Figs. 4f, 5e and 7b). The reversal of the superconducting diode polarity with varying temperatures rules out the possibility that the SDE in $CsV_3Sb_5$ flake devices were induced by the remnant magnetic field, since the remnant magnetic field, if exists, would be fixed and not induce the polarity change with temperatures. Furthermore, the field-free nature of the SDE in the flake devices is confirmed by the unchanged polarity under flipping the device at zero field (shown in Supplementary Fig. 6).

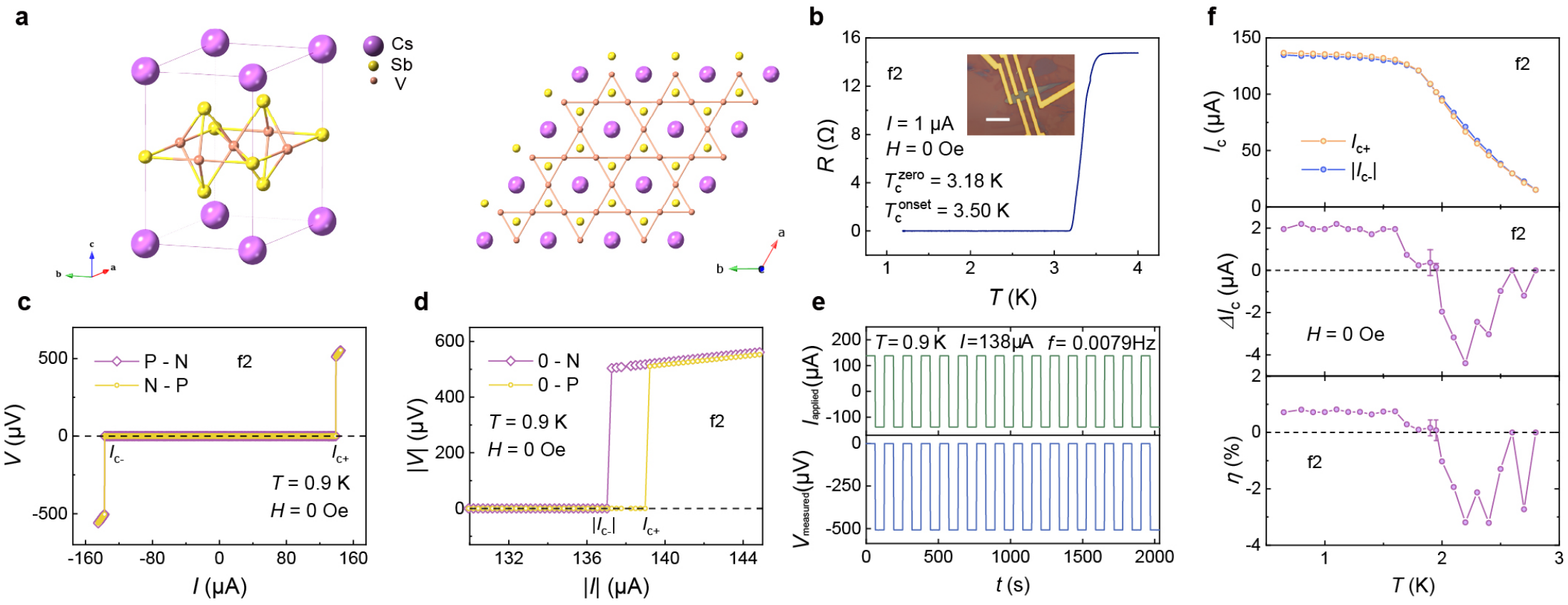

**Fig. 1 | Schematic crystal structure and the zero-field superconducting diode effect (SDE) in the $CsV_3Sb_5$ flake device f2. a**, Schematic crystal structure of $CsV_3Sb_5$ with purple, orange and yellow spheres denoting Cs, V and Sb atoms. **b**, Resistance ($R$)-temperature ($T$) curve of flake device f2 at zero magnetic field. The superconducting transition with the onset temperature $T_c^{onset}$ ~ 3.50 K and the zero-resistance temperature $T_c^{zero}$ ~3.18 K is observed. Inset shows the optical image of flake device f2. The scale bar represents 20 μm. **c**, $V$–$I$ curves of flake device f2 at 0.9 K and zero magnetic field, with the yellow curve denoting ramping the current from negative to positive and the purple curve denoting ramping the current from positive to negative. **d**, $V$–$I$ curves containing 0-P (current sweeping from zero to positive, yellow line) and 0-N (current sweeping from zero to negative, purple line) branches at 0.9 K and zero magnetic field, showing a clear difference (about 2 μA) between positive and negative critical currents ($I_{c+}$ and $|I_{c-}|$). **e**, Half-wave rectification of flake device f2 at 0.9 K and zero field. The top panel shows the applied d.c. current switching between +138 and −138 μA with a frequency of 0.0079 Hz. The bottom panel is the coincidentally measured voltage, showing an alternating switching between superconducting and resistive state depending on the direction of the current. **f**, $I_{c+}$ and $|I_{c-}|$ (the upper panel, orange dots and blue dots, respectively), $\Delta I_c$ (the middle panel, purple dots) and the diode efficiency $\eta$ ($\eta = \frac{I_{c+} - |I_{c-}|}{I_{c+} + |I_{c-}|} \times 100\%$ the lower panel, purple dots) as a function of temperature. The error bars in the middle and lower panels represent the standard deviation of $\Delta I_c$ and $\eta$, respectively, obtained from repeated measurements.

The observation of the field-free SDE indicates TRS breaking at zero field in our $CsV_3Sb_5$ flake devices. The loop current order[8–10,64–66] proposed in kagome superconductors offers a plausible explanation for the TRS breaking in $CsV_3Sb_5$. Specifically, the metallic CDW state with the TRS breaking loop-current order in $CsV_3Sb_5$ corresponds to a doped orbital Chern insulator, i.e. a Chern metal with a partially filled Chern band[8]. The pairing of the quasiparticles in the partially filled Chern band gives rise to a TRS breaking chiral topological superconductor, with chiral edge states carrying nonzero electrical currents along the chiral domain walls and the sample boundary[35], which are important for observing the nonreciprocal critical currents[66]. In a system that spontaneously breaks the $Z_2$ TRS, there are two kinds of domains with opposite chirality in the sample, which form stochastically below an onset temperature $T^*$ for TRS breaking[28,63,67]. As a result, the polarity of the SDE at low temperatures, which is determined by the distribution of the domains, depends on the thermal history across $T^*$. Theoretically, the onset temperature of TRS breaking ($T^*$) corresponds to the loop-current ordering temperature, which may be at or below the CDW formation temperature. To develop deeper insights into the nature of the field-free SDE and the polarity reversal behavior in $CsV_3Sb_5$ flake devices, we study the evolution of the SDE in flake devices f1-f4 under thermal cycles at zero magnetic field. As shown in Fig. 2a, in each cycle, the flake devices are first warmed up to the room temperature (well above $T_{CDW}$) at zero magnetic field, followed by decreasing the temperature into the zero-resistance state and measuring the $V$-$I$ curves. As shown in Figs. 2b-d and Supplementary Figs. 8-10, the polarity of the SDE in the $CsV_3Sb_5$ flake devices depends on thermal history and can be reversed after thermal cycling, indicating the existence of domains with TRS breaking in our $CsV_3Sb_5$ flake devices. Compared with f2 and f3, the SDE in the larger devices f1 and f4 exhibits a more pronounced dependence on thermal history (shown in Supplementary Figs. 9, 10). Our experimental results on the different devices further confirm that the devices with larger widths more frequently exhibit polarity reversal after thermal cycles (shown in Supplementary Tables 2, 3 and Supplementary Figs. 9-11, see Supplementary Note 1 for detailed discussion). Notably, the maximum diode efficiency is also tunable through zero-field thermal cycling, with the highest values exceeding 40% (Fig. 2e).

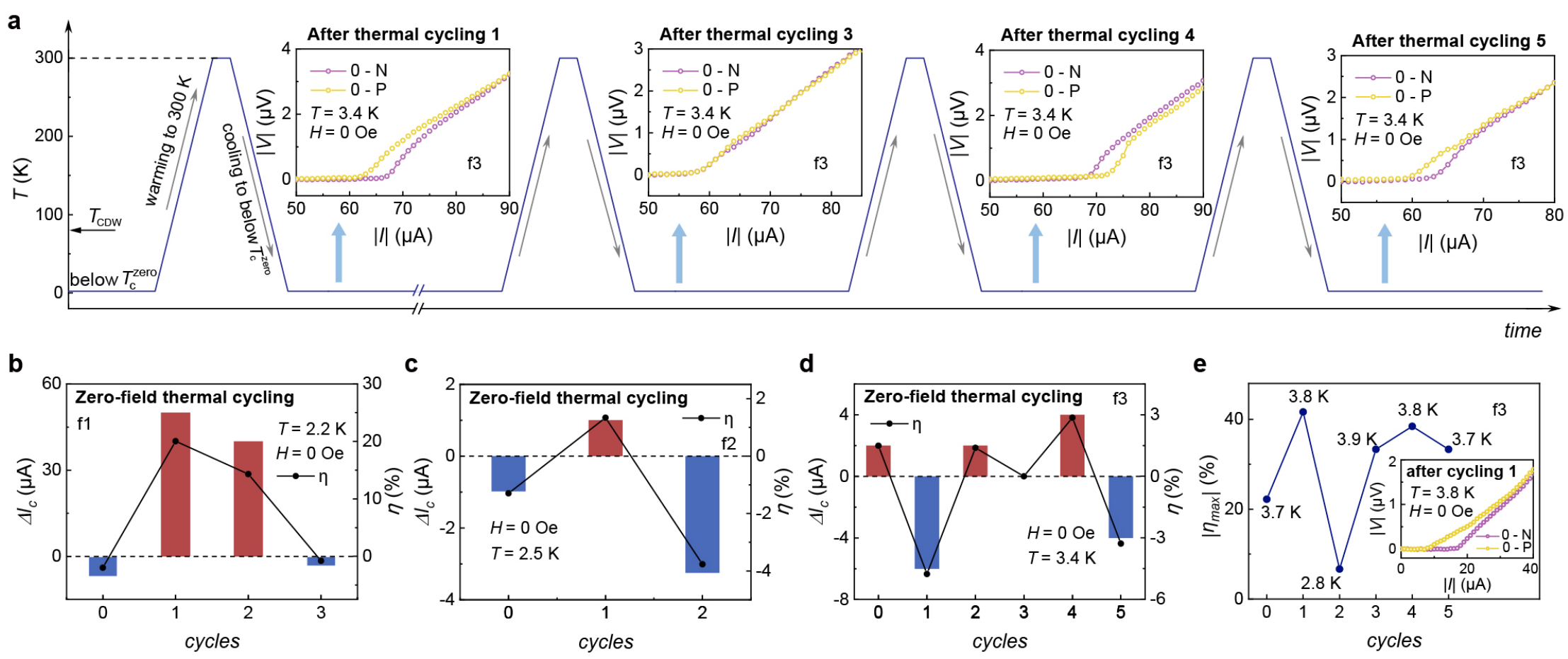

**Fig. 2 | Thermal modulation on the superconducting diode effect (SDE) in the $CsV_3Sb_5$ flake devices f1-f3. a,** Schematic illustration of the thermal cycling process. For each thermal cycle, the temperature is increased to 300 K (well above $T_{CDW}$) at zero magnetic field and then decreased to reach the zero-resistance superconducting state below $T_c^{zero}$ at zero magnetic field. Insets show the typical $V$–$I$ curves showing polarity reversal under thermal cycling 1, 3, 4, and 5 in f3 at 3.4 K. **b-d**, $\Delta I_c$ (the red and blue bars, left axis) and diode efficiency (the black line chart, right axis) of f1 at 2.2 K (b), f2 at 2.5 K (c) and f3 at 3.4 K (d) as a function of cycle numbers. Here, the color and height of the bars represent the sign (red for positive and blue for negative) and the amplitude of $\Delta I_c$, respectively. The sign change in $\Delta I_c$ under thermal cycles can

be clearly observed. **e**, The absolute value of the maximum diode efficiency $|\eta_{max}|$ as a function of thermal cycling numbers in f3. As shown in Supplementary Fig. 5e, the diode efficiency $\eta$ ($\eta = \frac{I_{c+} - |I_{c-}|}{I_{c+} + |I_{c-}|} \times 100\%$) depends on temperature and reaches a maximum value at a specific temperature. After the first cycle, the maximum diode efficiency is observed at 3.8 K, which exceeds 40%. The inset shows the corresponding *V–I* curves measured at 3.8 K after thermal cycling 1.

**Nonreciprocal superconducting critical currents with normal state field trainability in $CsV_3Sb_5$ thin flake devices**

Physically, for a state with spontaneous TRS breaking and chiral domains, the cooling through its onset temperature ($T^*$) in the presence of a magnetic field tends to align the domains, making the direction of TRS breaking trainable. To get deeper insights into the mechanism of the observed TRS breaking in the superconducting state, we perform systematic field-training measurements on the $CsV_3Sb_5$ flake devices. Figure 3 presents the results for flake device f4. As illustrated in Fig. 3a, c, for each training cycle, a magnetic field of ±100 Oe is first applied at 300 K. The sample is subsequently cooled to a target temperature—either 5 K (Fig. 3a, b) or 50 K (Fig. 3c, d) in separate cooling cycles. For each temperature condition, multiple independent training cycles (11 cycles in Fig. 3a, b and 8 cycles in Fig. 3c, d) are performed. Both temperatures are situated below the CDW transition but strictly above the superconducting transition. At these respective temperatures, the magnetic field is carefully reduced to zero, verified by a high-precision Hall sensor (see Methods). The sample is then further cooled to the zero-resistance state to measure the *V-I* curves. Figure 3b displays the $\Delta I_c$ at 2.0 K obtained after removing the training field at 5 K. Evidently, based on the statistics from these repeated 11 cycles, the SDE is trainable: the polarity is consistently negative under positive-field cooling and positive under negative-field cooling. Notably, the SDE polarity established by the field training process remains invariant with temperature variations, as shown in Supplementary Fig. 12. Figure 3d displays the $\Delta I_c$ at 2.0 K obtained after removing the training field at 50 K. The SDE is trainable too.

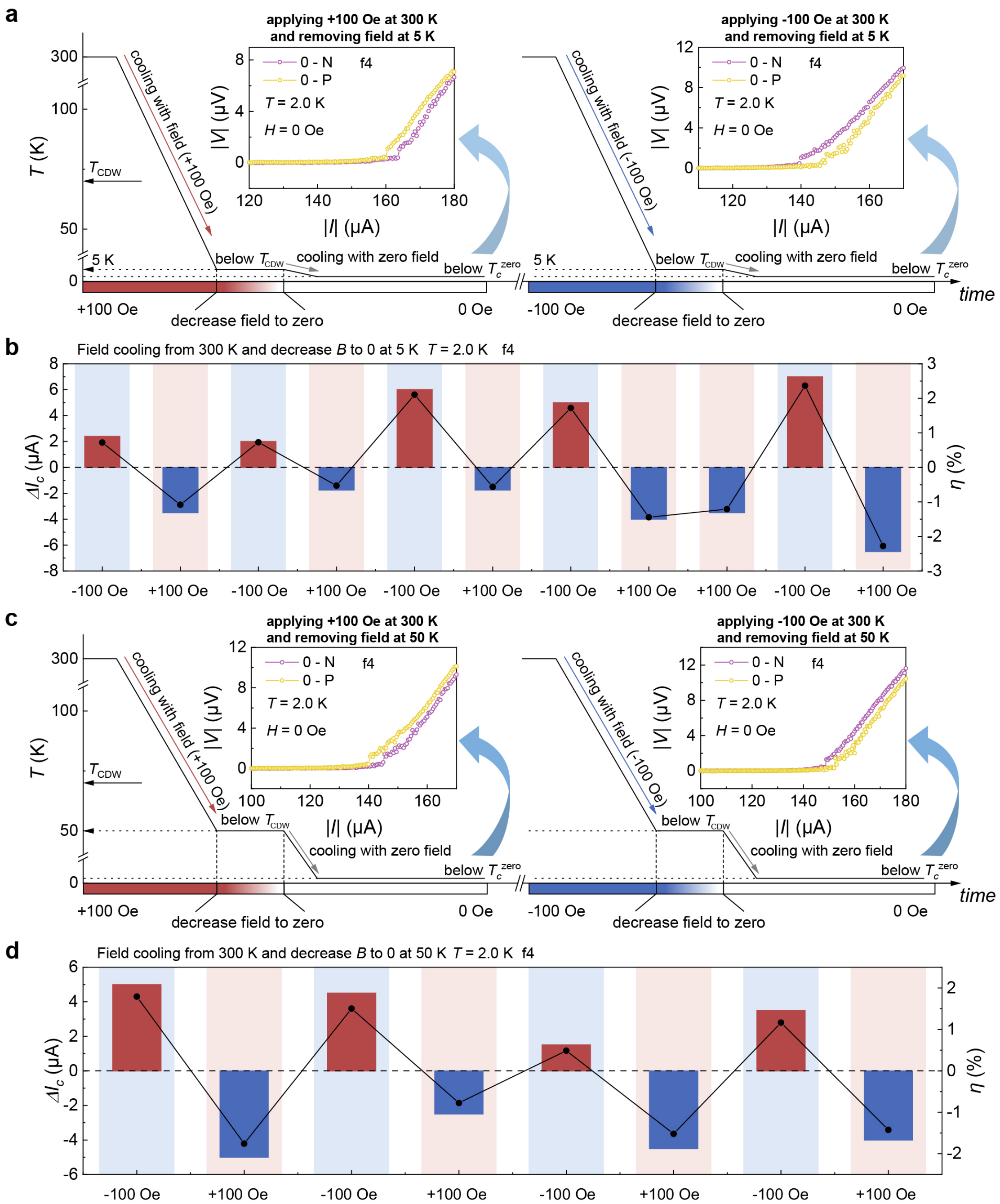

**Fig. 3 | The superconducting diode effect (SDE) with normal state field trainability in kagome superconductor $CsV_3Sb_5$ flake devices.** Field training on the SDE in the $CsV_3Sb_5$ flake device f4, with the magnetic field removed at 5 K (a, b) and 50 K (c, d). **a, c,** Schematic illustration of the field training process. The left panel shows the procedure under a +100 Oe (indicated in red on the horizontal axis) field training, while the right panel corresponds to -100 Oe (indicated in blue on the horizontal axis) field training. During field training, a magnetic field (+100 Oe or -100 Oe) is first applied at room temperature, followed by cooling the sample down to 5 K or 50 K under the field. These temperatures are below the charge density wave (CDW) transition temperature ($T_{CDW}$~70 K) of device f4. The magnetic field is then decreased to zero at 5 K or 50 K, and the sample is further cooled to below the zero-resistance temperature, where the *V*-

$I$ measurements are performed. Insets show the $V$-$I$ curves measured at 2 K after +100 Oe (left panel) and -100 Oe (right panel) field training. **b, d,** The polarity change at 2.0 K in f4 generated by reversing the field in the normal state. The color and height of the bars represent the sign (dark red for positive, dark blue for negative) and the amplitude of $\Delta I_c$, respectively. The background shading indicates the direction of training field (light red for +100 Oe and light blue for -100 Oe). The black dots correspond to the diode efficiency $\eta$ (right axis).

Crucially, although the high-precision Hall sensor already confirms the zero-field condition, to further rigorously exclude the possible influence of the magnet's remnant field, we conduct *in-situ* sample flipping experiments. The SDE polarity remains unchanged after flipping the device, confirming that the training effect is intrinsic (see Supplementary Fig. 13). Furthermore, the trainable SDE is also observed in other flake devices (f5, f3 and f6) with field removal below $T_{CDW}$, demonstrating the universality and robustness of our observations (see Supplementary Figs. 14-16, Supplementary Table 4). Additionally, as shown in Supplementary Fig. 17, control experiments performed within the CDW phase (i.e. both magnetic field application and removal below $T_{CDW}$) demonstrate that the SDE polarity can also be trained by a magnetic field within the CDW phase, serving as further evidence for the TRS breaking in the CDW phase.

To further verify the origin of this trainability, we perform a control experiment by removing the field at 150 K (well above $T_{CDW}$). In this case, the SDE polarity is not dependent on the direction of the applied magnetic field and becomes stochastic (Fig. 4), which implies that the TRS breaking in the superconducting state is inherited from the CDW state. Indeed, in the loop current CDW scenario, reversing the direction of the loop currents in the metallic CDW state changes the direction of the circulating loop supercurrents as well as the chirality of both the TRS breaking superconducting state and the associated chiral edge states[35], thereby reversing the polarity of the SDE, which is consistent with the polarity reversal of the SDE in our experimental results.

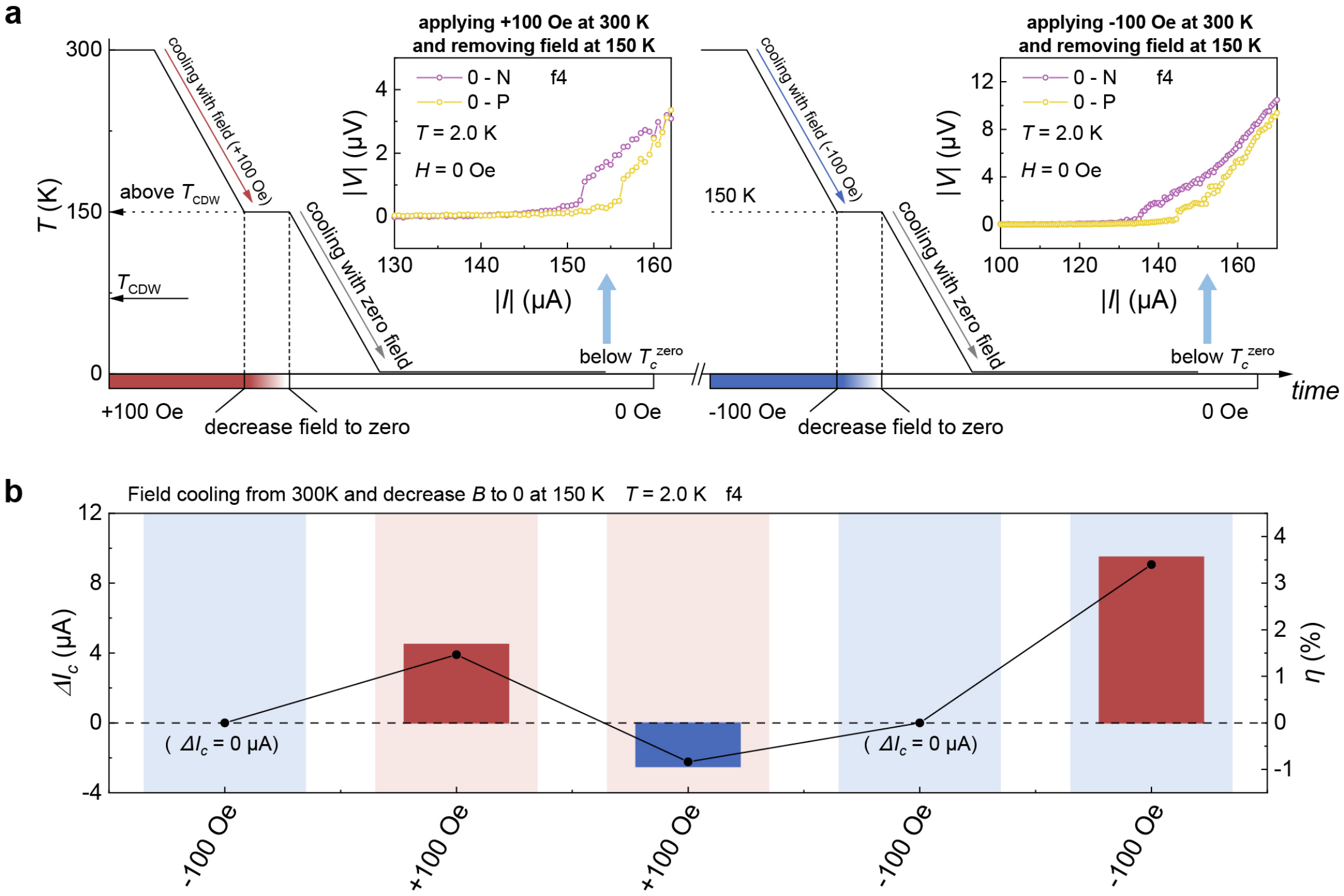

**Fig. 4 | Field training on the superconducting diode effect (SDE) in the $CsV_3Sb_5$ flake device f4, with the magnetic field removed at 150 K. a**, Schematic illustration of the field training process. The left panel shows the procedure under a +100 Oe (indicated in red on the horizontal axis) field training, while the right panel corresponds to -100 Oe (indicated in blue on the horizontal axis) field training. During field training, a magnetic field (+100 Oe or -100 Oe) is first applied at room temperature, followed by cooling the sample down to 150 K under the field. This temperature is above the charge density wave (CDW) transition temperature ($T_{CDW}$~70 K) of device f4. The magnetic field is then decreased to zero at 150 K, and the sample is further cooled to below the zero-resistance temperature, where the *V-I* measurements are performed. Insets show the *V-I* curves measured at 2 K after +100 Oe (left panel) and -100 Oe (right panel) field training. **b**, The SDE at 2.0 K in f4. The color and height of the bars represent the sign (dark red for positive, dark blue for negative) and the amplitude of $\Delta I_c$, respectively. The background shading indicates the direction of training field (light red for +100 Oe and light blue for -100 Oe). The black dots correspond to the diode efficiency $\eta$ (right axis).

The observation of a zero-field SDE is typically interpreted as a signature of TRS breaking in the superconducting state[28,40,41,43,63]. Our work advances this understanding by identifying the origin of the symmetry breaking via field training. The unique ability and the training protocol to modulate the polarity of the SDE by applying a field in the normal state provides compelling evidence that the TRS breaking order parameter is already established in the normal state below the CDW transition, long before the system enters the superconducting phase. It implies that the TRS-breaking domains exist in the normal state and serve as a template for the superconducting state. This "memory" effect, where information is written in the normal phase and read out in the superconducting phase, makes the kagome system a promising platform to study novel superconductivity and SDE.

**Zero-field SDE in $CsV_3Sb_5$ micro-bridge devices**

The results from the flake devices provide valuable insights into the TRS breaking in $CsV_3Sb_5$ superconductors. To achieve better control over the SDE, we further fabricate $CsV_3Sb_5$ micro-bridge devices (*R*-*T* curves shown in Supplementary Fig. 18). The $CsV_3Sb_5$ bridge devices are fabricated by etching the kagome superconductor thin flakes exfoliated from bulk samples, as shown in the inset of Fig. 5a. The SDE at zero magnetic field and the stable rectification effect are observed in micro-bridge devices s1 with a length of around 2.83 µm and a width of 479 nm (Fig. 5a, b) and s2 with a length of around 2.80 µm and a width of 516 nm (Fig. 5c, d). The field-free nature of the SDE in the $CsV_3Sb_5$ micro-bridge device is demonstrated by flipping the device at zero magnetic field. Figure 5e shows the SDE in s2 at 0.5 K, and zero magnetic field, with the device facing up (the schematic position of the device shown in the inset of Fig. 5e). When flipping the device (the schematic position of the device shown in the inset of Fig. 5f) at 0.5 K and zero magnetic field, the polarity of SDE (shown in Fig. 5f) remains the same as that in Fig. 5e, demonstrating that the SDE is not induced by the possible remnant magnetic field in the measurement system, confirming the field-free nature of the SDE.

Figure 5g, h displays the SDE as a function of temperature in the micro-bridge device s2. With increasing temperature, $\Delta I_c$ decreases, while the sign of $\Delta I_c$ is always positive (Fig. 5h), in contrast to the polarity reversal behavior observed with increasing temperature in the flake devices (Fig. 1f, Supplementary Fig. 4f, 5e, 7b). Furthermore, the SDE polarity in the micro-bridges is insensitive to thermal cycling (Supplementary Fig. 11). These results suggest that the thermal fluctuations of domains, which influence the polarity reversal in the flake devices, are strongly suppressed in the micro-bridge devices with the small widths (Supplementary Table 3). The latter may arise from the pinning effect induced by disorder from the etching process or the reduced area of the micro-bridge device.

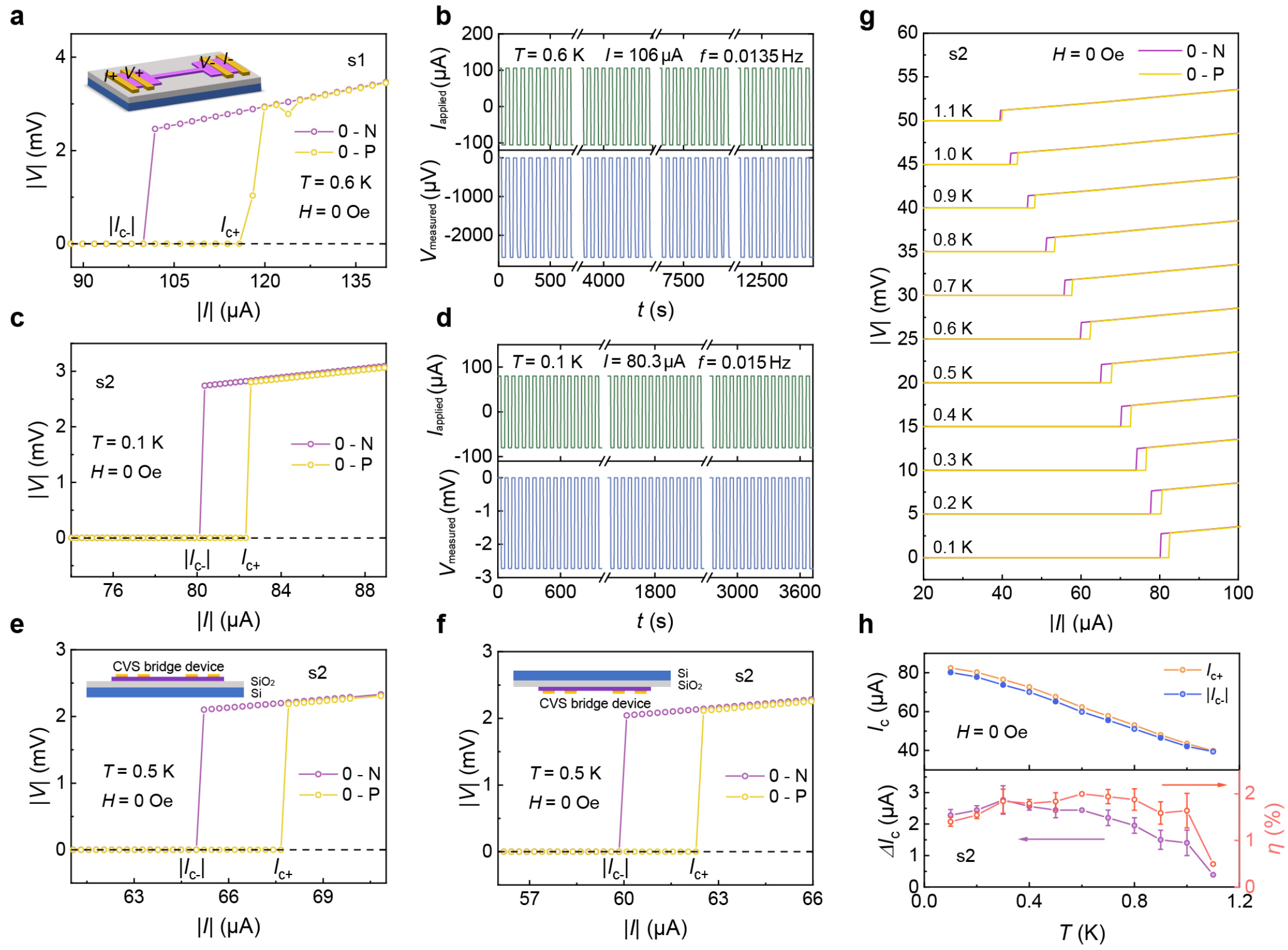

**Fig. 5 | The zero-field SDE in $CsV_3Sb_5$ micro-bridge devices. a**, *V–I* curves containing 0-P and 0-N branches in micro-bridge device s1 at 0.6 K and zero magnetic field, showing a clear difference (about 16 μA) between positive and negative critical currents ($I_{c+}$ and $|I_{c-}|$). **b**, Half-wave rectification of s1 at 0.6 K and zero field. The top panel shows the applied d.c. current switching between +106 and –106 μA with a frequency of 0.0135 Hz. The bottom panel is the coincidentally measured voltage, showing an alternating switching between superconducting and resistive state depending on the direction of the current. **c**, *V–I* curves containing 0-P and 0-N branches in micro-bridge device s2 at 0.1 K and zero magnetic field. **d**, Half-wave rectification of s2 at 0.1 K and zero field. The amplitude of the excitation current is 80.3 μA and the frequency is 0.015 Hz. **e**, **f**, The zero-field SDE in micro-bridge device s2 before (**e**) and after flipping (**f**) the device. The SDE polarity shown in **f** is the same as that before reversing the device (shown in **e**), confirming the field-free nature of SDE. Note that $I_{c+}$ and $|I_{c-}|$ in **f** are slightly smaller than that in **e**, which may result from the degradation of the device during *ex situ* flipping the device at room temperature. **g**, The absolute *V-I* values in 0-P (yellow lines) and 0-N (purple lines) branches at various temperatures in s2. **h**, $I_{c+}$ (the upper panel, orange dots), $|I_{c-}|$ (the upper panel, blue dots), $\Delta I_c$ (the lower panel, purple dots) and the diode efficiency $\eta$ ($\eta = \frac{I_{c+} - |I_{c-}|}{I_{c+} + |I_{c-}|} \times 100\%$ the lower panel, red dots) as a function of temperature in s2. The error bars in the lower panel represent the standard deviation of $\Delta I_c$ and $\eta$ within repeated measurements.

In summary, we observe the zero-field nonreciprocal critical currents, i.e. the field-free

SDE, in $CsV_3Sb_5$ flake and micro-bridge devices, indicating TRS breaking in kagome superconductor $CsV_3Sb_5$. The polarity of the SDE can be reversed when changing the direction of TRS breaking in the normal state with CDW orders by the magnetic training, revealing the intertwined relation between the TRS breaking CDW state and the chirality of the superconducting state in $CsV_3Sb_5$. Our observations provide new insights and will inspire future investigations into the mechanism for novel spontaneous TRS breaking in kagome metals and superconductors.

## Methods

### Crystal growth

We used the self-flux method to grow $CsV_3Sb_5$ single crystals[11]. Firstly, the mixture of Cs ingot (purity 99.75%), V powder (purity 99.9%) and Sb grains (purity 99.999%) was put into an alumina crucible and sealed in a quartz ampoule under argon atmosphere. Then, heated up the quartz ampoule to 1273 K for 12 h and held for 24 h. Subsequently, rapidly cooled down the quartz ampoule to 1173 K in two hours and slowly cooled down the quartz ampoule to 923 K. Finally, the $CsV_3Sb_5$ single crystals were separated from the flux by using a centrifuge. In order to prevent the reaction of Cs with air and water, all the preparation processes except the sealing and heat treatment procedures were carried out in an argon-filled glove box.

### Devices fabrication

The thin $CsV_3Sb_5$ flakes were firstly exfoliated from bulk single crystals using the scotch tape in an argon-filled glove box with the $O_2$ and $H_2O$ levels below 0.1 ppm, and then transferred onto the 300-nm-thick $SiO_2$/Si substrates, which were pre-cleaned in oxygen plasma for 5 min at ~80 mTorr pressure. For flake devices f1-f3 and f5-f7, electrodes were patterned by a standard electron beam lithography (EBL) process in a FEI Helios NanoLab 600i Dual Beam System with PMMA 495A11 as a resist, and metal electrodes (Ti/Au, 6.5/180 nm) were deposited in a LJUHVE-400 L E-Beam Evaporator. For f4, the EBL process was conducted using a TESCAN MIRA SEM, and pure Au (130 nm) were deposited using a DE 400 system. Note that the PMMA 495A11 resist was not baked. Finally, the PMMA layers were removed by standard lift-off process and $CsV_3Sb_5$ flake devices with four or six metal electrodes were obtained.

For the micro-bridge devices s1 and s2, PMMA 495A11 electron-beam resist (800 nm thick, without baking) were spin-coated on the above-obtained flake devices as the protection layers and focused ion beam (FIB) was used to etch micro-bridge structures on $CsV_3Sb_5$ flake devices in an FEI Helios NanoLab 600i Dual Beam System[11]. The beam current for etching was 7.7 pA.

### Transport measurements

Standard four-electrode method was used to characterize the transport properties of $CsV_3Sb_5$ flake and micro-bridge devices. The measurements for flake devices f1, f2, and micro-bridge device s2 (via a dilution refrigerator option), s1 (via a $^3$He option), and f4–f7 (via a rotator option) were all conducted in a Re-liquefier based 16

T Physical Property Measurement System (PPMS-16, Quantum Design). The transport measurements of flake device f3 were conducted with a dilution refrigerator option in the cryogen-free physical property measurement system (PPMS Dynacool 14 T). As For the *V-I* curves and half-wave rectification measurements, for devices f3–f7**,** a Keithley 6221 AC/DC Current Source Meter was used to apply the current (d.c. or square-wave), and the voltage was measured by a Keithley 2182A Nanovoltmeter. The other devices (f1, f2, s1, s2) were measured using the built-in d.c. mode of the PPMS. The applied DC current on f1-f7 and s1, s2 was automatically cut off between two adjacent voltage measurements to reduce the possible heating effect in the measurements.

For the field-training experiments, a precise compensation protocol was implemented to achieve a zero-field condition after the removal of the training field. For example, in the -100 Oe training experiments on flake devices f4–f7, the magnetic field was firstly ramped to -5 Oe at a rate of 1 Oe/s. Then, the magnetic field was further slowly ramped at a rate of 0.1 Oe/s to a target setpoint of +2.8 Oe (the empirical zero-field setpoint for this field ramping procedure in our PPMS system). At the target stage, a Hall sensor (Cryogenic transverse Hall sensor HGCT 3020, Lakeshore) indicated that the actual magnetic field was typically ~ 0.1 Oe. A final fine-tuning compensating field (typically ~ -0.1 Oe) was applied to further suppress the residual magnetic field to below 0.01 Oe monitored by the Hall sensor.

**AFM measurements**

The AFM measurements of the $CsV_3Sb_5$ flakes and micro-bridge devices were performed in the Bruker DimensionICON system. The thicknesses of flake devices f1-f7 and micro-bridge devices s1, s2 are 23 nm, 8.8 nm, 33 nm, 22.5 nm, 33 nm, 15.5 nm, 25 nm, 27.5 nm and 20 nm, respectively.

**Data availability**

Relevant data supporting the key findings of this study are available within the article and the Supplementary Information file. All raw data generated during the current study are available from the corresponding author upon request.

## Acknowledgements

We thank Chandra Varma for insightful discussions and suggestions regarding the field-training experiments, and acknowledge technical assistance from Chunsheng Gong and Zhijun Tu. J.W. acknowledges the support of the National Natural Science Foundation of China (12488201), and the Quantum Science and Technology-National Science and Technology Major Project (2021ZD0302403). J.G. acknowledges the support of the National Natural Science Foundation of China (12504158, 92565201). H.L. acknowledges the support of the National Natural Science Foundation of China

(12274459), and the National Key R&D Program of China (2022YFA1403800, 2023YFA1406500). P.W. acknowledges the support of the National Natural Science Foundation of China (124B2067). H.P. acknowledges the support of the National Natural Science Foundation of China (124B1036). Z.W. acknowledges the support of the U.S. Department of Energy, Basic Energy Sciences Grant DE-FG02-99ER45747 and Research Corporation for Science Advancement under Cottrell SEED Award No. 27856.

**Author contributions**

J.W. conceived and instructed the research. J.G. and X.L. performed the transport measurements. J.G., X.L. and H.P. analyzed the data under the guidance of J.W.. Z.W. contributed to the theoretical explanation. Q.Y. and H.L. grew the single crystals. J.G. X.L. and P.W. fabricated the devices. J.G., X.L., Z.W. and J.W. wrote the manuscript with the input from P.W. and other authors.

**Competing interests**

The authors declare no competing interests.

## Supplementary Information

**Supplementary Table 1 | The contact resistance and four-terminal resistance (sample resistance) in flake devices f1-f7.**

| | Four-terminal resistance (Ω) | Contact resistance (Ω) |
|---|---|---|
| f1 | 2.44 (5 K) | 4.57 (5 K) |
| f2 | 18.85 (5 K) | 9.20 (5 K) |
| f3 | 17.79 (300 K) | 20.66 (300 K) |
| f4 | 34.46 (300 K) | 31.72 (300 K) |
| f5 | 0.56 (4.5 K) | 1.52 (2 K) |
| f6 | 71.98 (300 K) | 8.03 (300 K) |
| f7 | 3.30 (5 K) | 3.47 (5 K) |

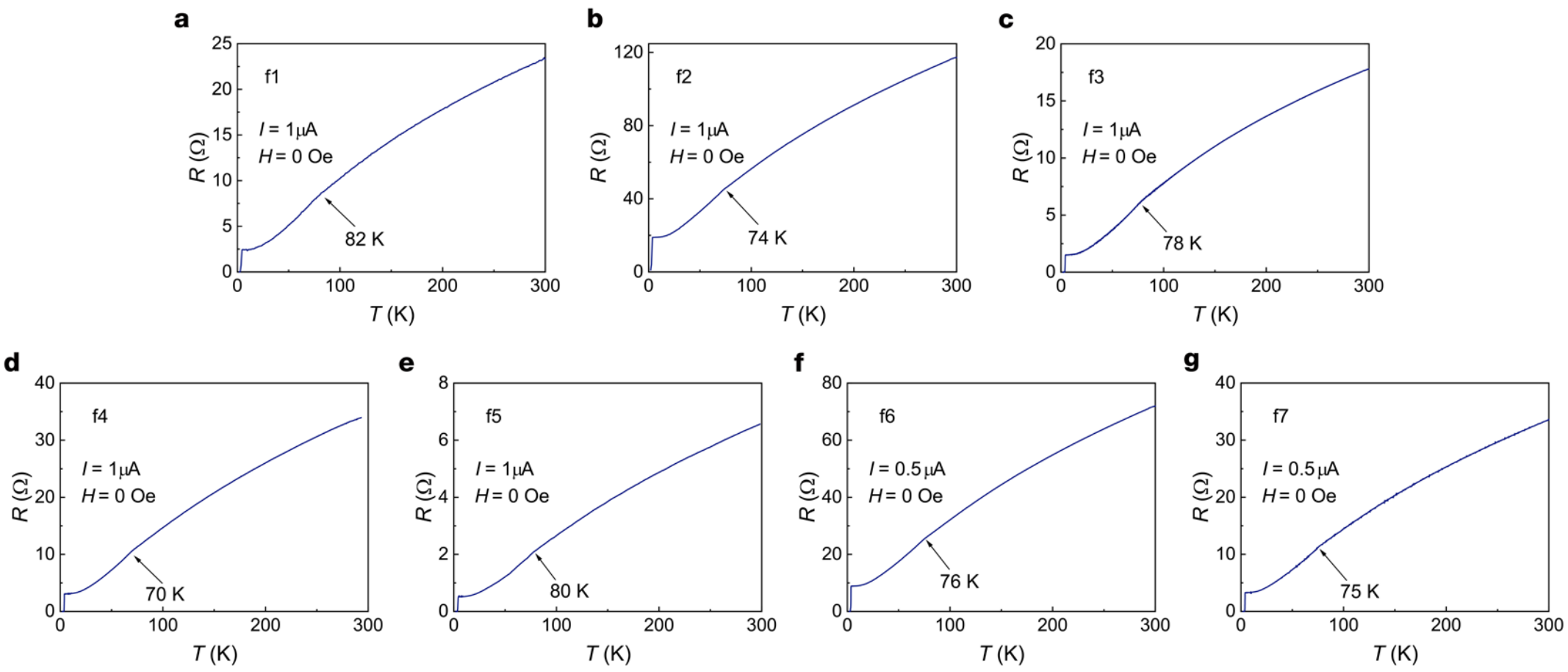

**Supplementary Fig. 1 | Resistance (*R*) as a function of temperature (*T*) in $CsV_3Sb_5$ flake devices f1-f7 at zero magnetic field.** The black arrows label the kinks in the *R*-*T* curves.

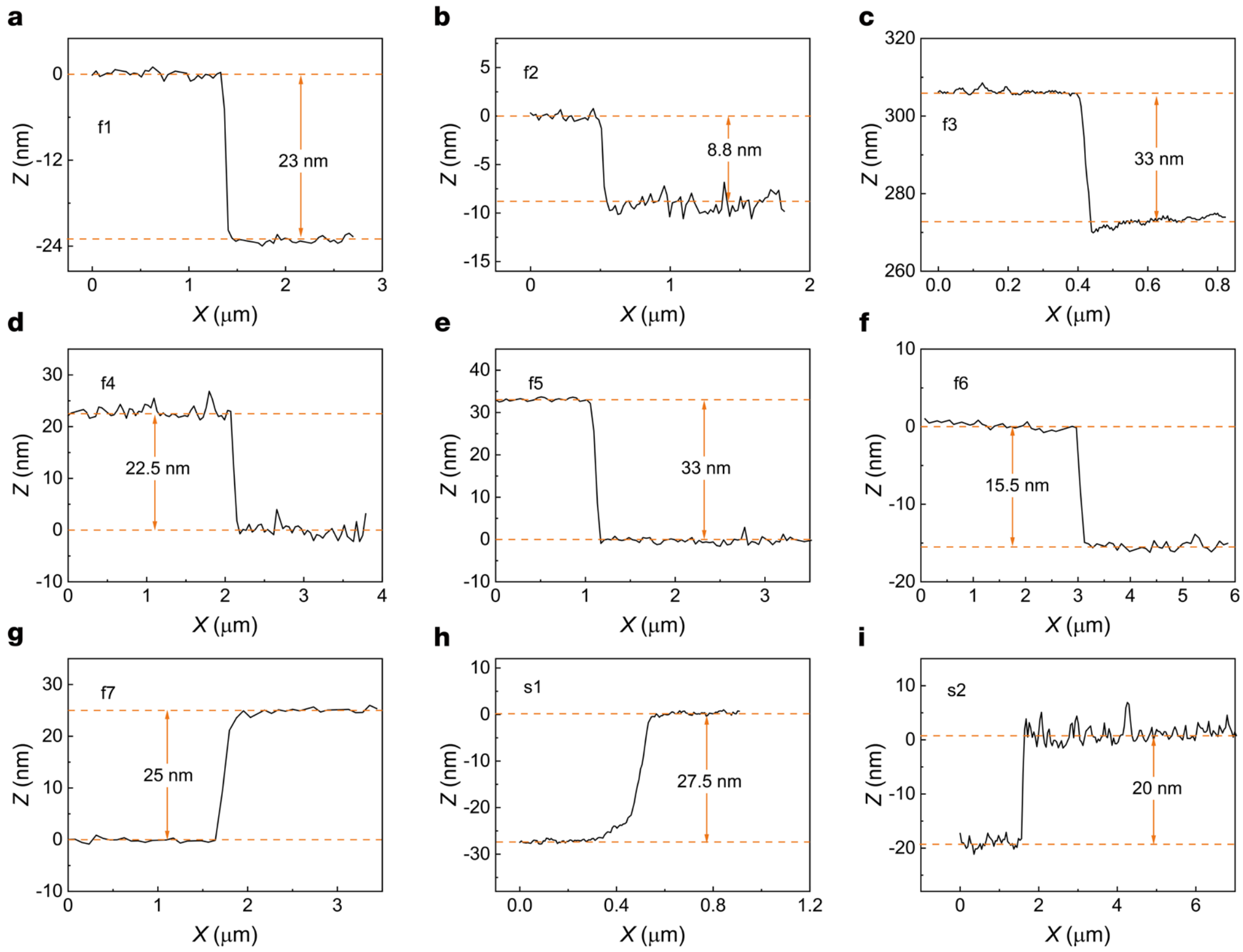

**Supplementary Fig. 2 | Thickness of $CsV_3Sb_5$ flake devices f1-f7 (a-g), micro-bridge devices s1 (h) and s2 (i) measured by atomic force microscopy.**

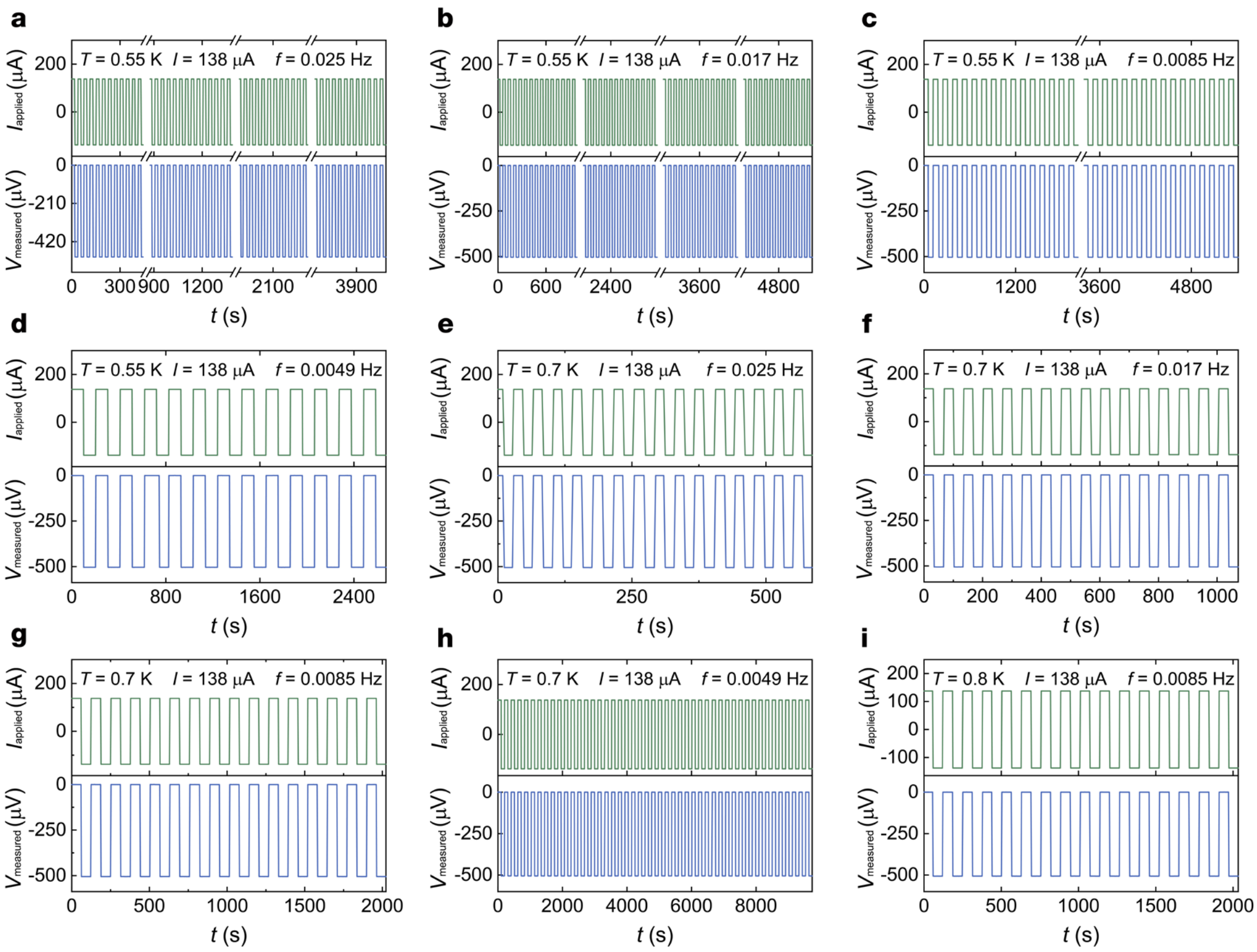

**Supplementary Fig. 3 | Stable half-wave rectification at various temperatures and**

**frequencies in $CsV_3Sb_5$ flake device f2. a-d**, Half-wave rectification at 0.55 K and zero field, with the applied current amplitude of 138 μA and frequency of 0.025 (a), 0.0017 (b), 0.0085 (c) and 0.0049 Hz (d). The top panel shows the applied d.c. current switching between +138 and −138 μA. The bottom panel is the coincidentally measured voltage, showing an alternating switching between superconducting and resistive state depending on the direction of the current. **e-h**, Half-wave rectification at 0.7 K and zero field, with the applied current amplitude of 138 μA and frequency of 0.025 (e), 0.0017 (f), 0.0085 (g) and 0.0049 Hz (h). **i**, Half-wave rectification at 0.8 K and zero field, with the applied current amplitude of 138 μA and frequency of 0.0085 Hz.

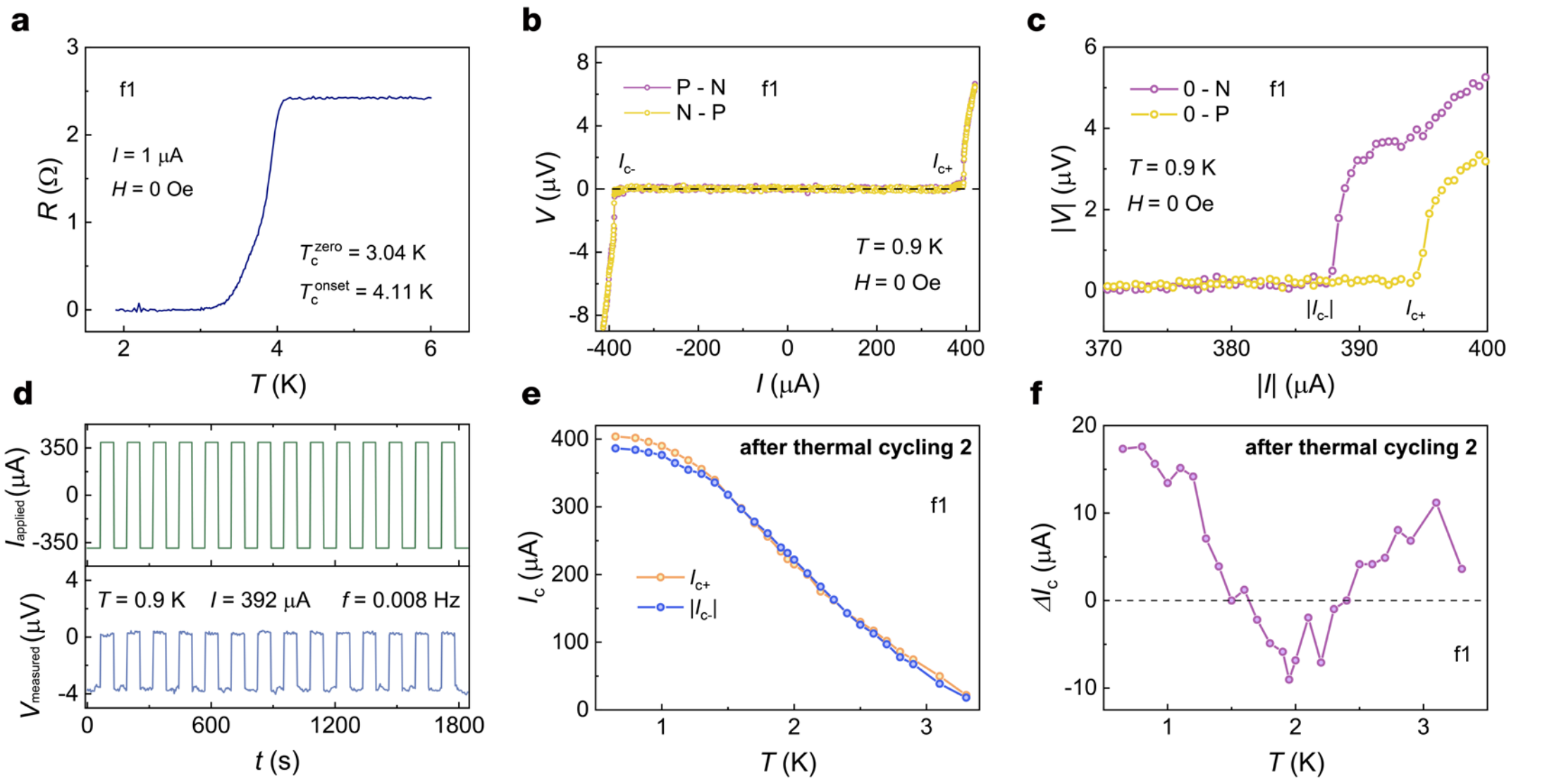

**Supplementary Fig. 4 | The zero-field superconducting diode effect (SDE) in the $CsV_3Sb_5$ flake device f1.** The length between two voltage electrodes, the width and thickness of f1 are 4.6 μm, 15.3 μm and 23 nm, respectively. **a**, Resistance (*R*)-temperature (*T*) curve of flake device f1 at zero magnetic field. The superconducting transition with the onset temperature $T_c^{onset}$ ~ 4.11 K and the zero-resistance temperature $T_c^{zero}$ ~3.04 K is observed. **b,** *V–I* curves of f1 at 0.9 K and zero magnetic field, with the yellow curve denoting positive sweep (ramping the current from negative to positive) and the purple curve denoting negative sweep (ramping the current from positive to negative). **c**, *V–I* curves containing 0-P (current sweeping from zero to positive, yellow line) and 0-N (current sweeping from zero to negative, purple line) branches at 0.9 K and zero magnetic field, showing a clear difference between positive and negative critical currents ($I_{c+}$ and $|I_{c-}|$). **d**, Half-wave rectification at 0.9 K and zero field, with the applied current amplitude of 392 μA and frequency of 0.008 Hz. **e**, $I_{c+}$ (orange dots) and $|I_{c-}|$ (blue dots) as a function of temperature. **f**, $\Delta I_c$ (purple dots) as a function of temperature. With increasing temperature, the sign of $\Delta I_c$ changes.

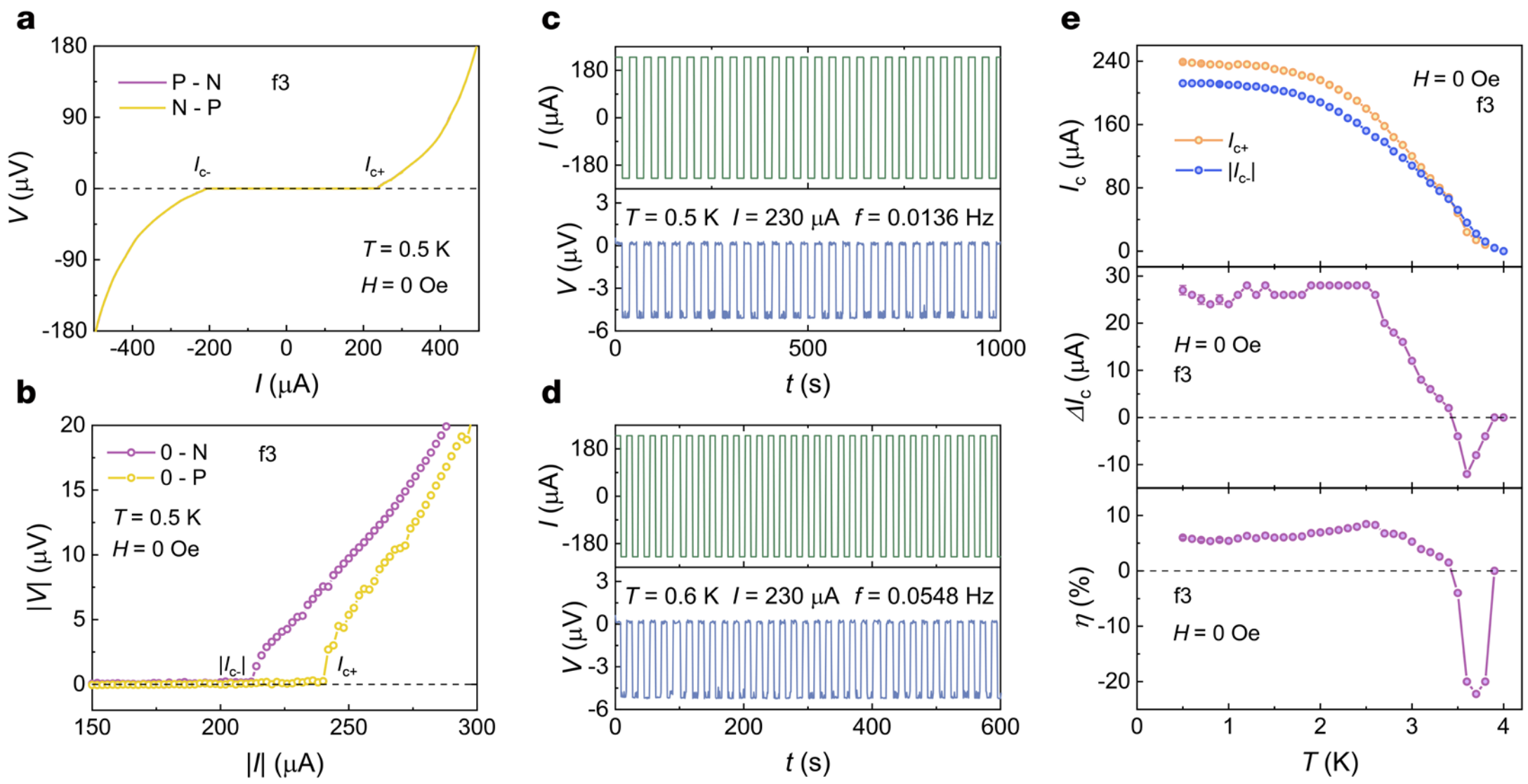

**Supplementary Fig. 5 | The zero-field superconducting diode effect (SDE) in the $CsV_3Sb_5$ flake device f3. a**, *V–I* curves containing N-P (current sweeping from negative to positive, yellow line) and P-N (current sweeping from positive to negative, purple line) branches at 0.5 K and zero magnetic field. **b**, *V–I* curves containing 0-P (current sweeping from zero to positive, yellow line) and 0-N (current sweeping from zero to negative, purple line) branches at 0.5 K and zero magnetic field, showing a clear difference (about 27 μA) between positive and negative critical currents ($I_{c+}$ and $|I_{c-}|$). **c**, Half-wave rectification of flake device f3 at 0.5 K and zero magnetic field. The top panel shows the applied d.c. current switching between +230 and –230 μA with a frequency of 0.0136 Hz. The bottom panel is the coincidentally measured voltage, showing an alternating switching between superconducting and resistive state depending on the direction of the current. **d**, Half-wave rectification of flake device f3 at 0.6 K and zero magnetic field. **e**, $I_{c+}$ (the upper panel, orange dots), $|I_{c-}|$ (the upper panel, blue dots), $\Delta I_c$ (the middle panel, purple dots) and the diode efficiency $\eta$ ($\eta = \frac{I_{c+} - |I_{c-}|}{I_{c+} + |I_{c-}|} \times 100\%$ the lower panel, purple dots) as a function of temperature. With increasing temperature, the sign of $\Delta I_c$ changes from positive to negative. A maximum diode efficiency of around 22% is detected at 3.7 K.

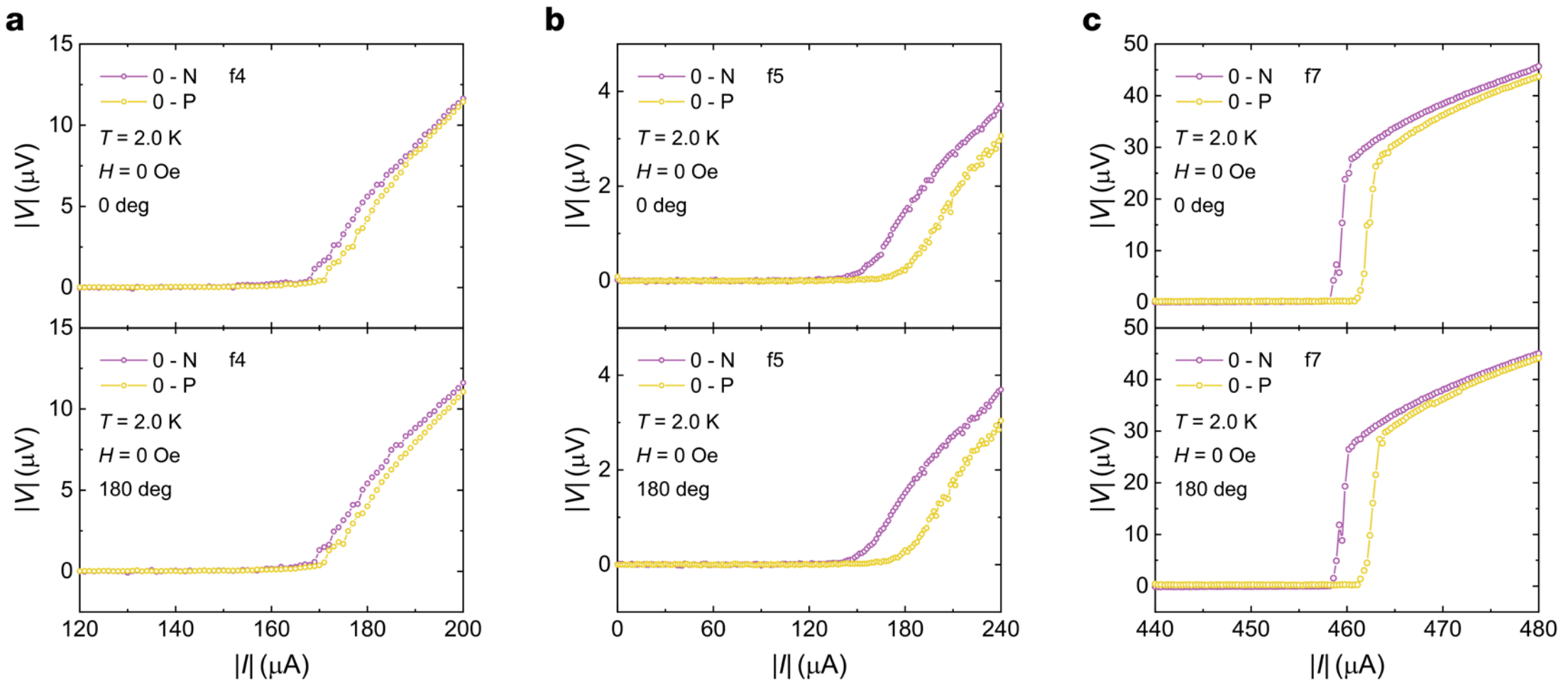

**Supplementary Fig. 6 | The zero-field SDE in the new flake devices f4 (a), f5 (b) and f7 (c) before (0 deg, the upper panel) and after flipping (180 deg, the lower panel) the device.** The polarity of SDE is the same as that before reversing the device, confirming the field-free nature of SDE in the $CsV_3Sb_5$ flake devices.

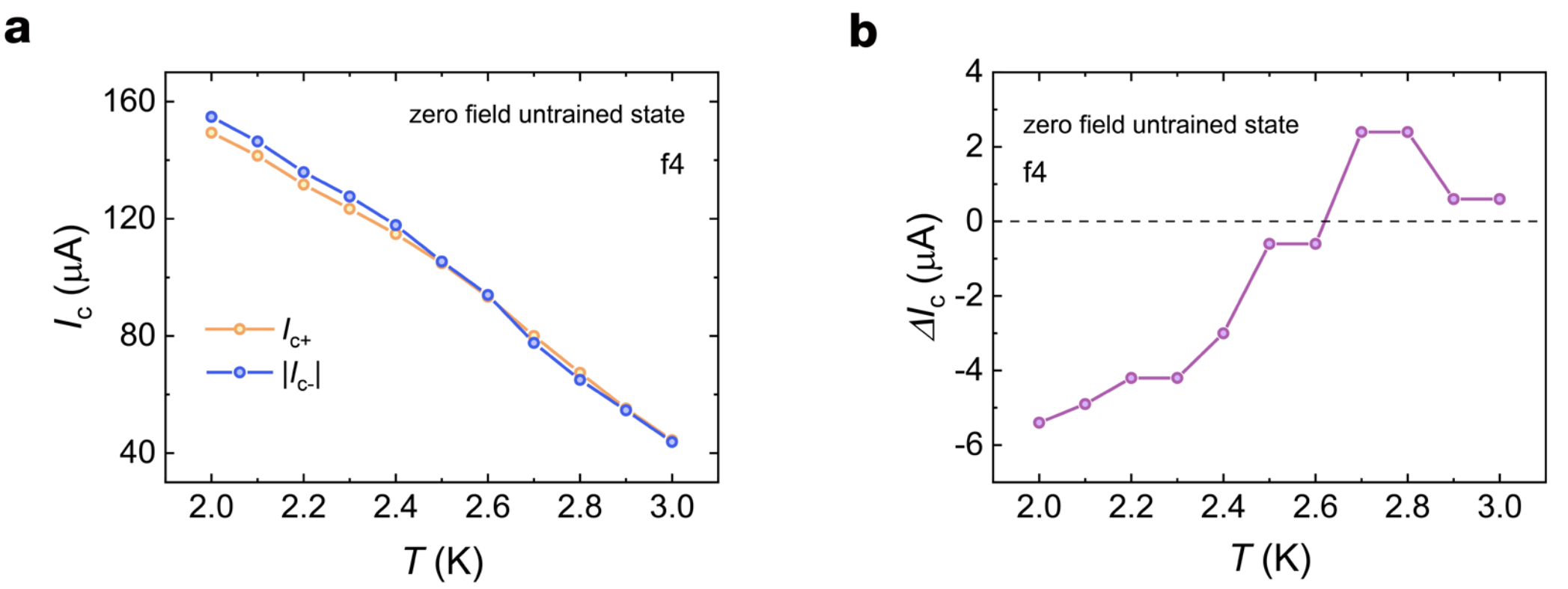

**Supplementary Fig. 7 | $I_{c+}$ (orange dots in a), $|I_{c-}|$ (blue dots in a) and $\Delta I_c$ (b) in the flake device f4 as a function of temperature, measured at zero magnetic field in the untrained state.** Here, the untrained state refers to the condition where the sample is cooled from 300 K in the absence of an external magnetic field (zero-field cooling).

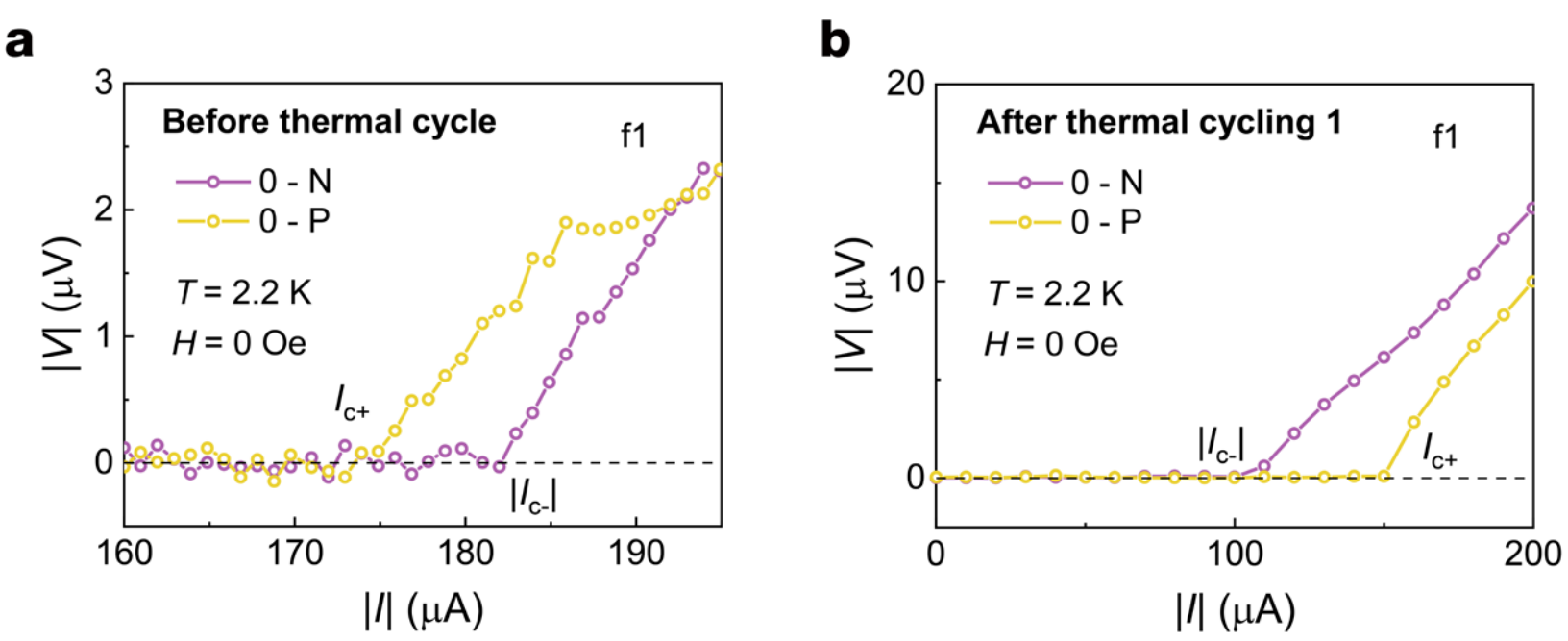

**Supplementary Fig. 8 | The superconducting diode effect (SDE) before (a) and after (b) zero-field thermal cycling 1 in flake device f1.**

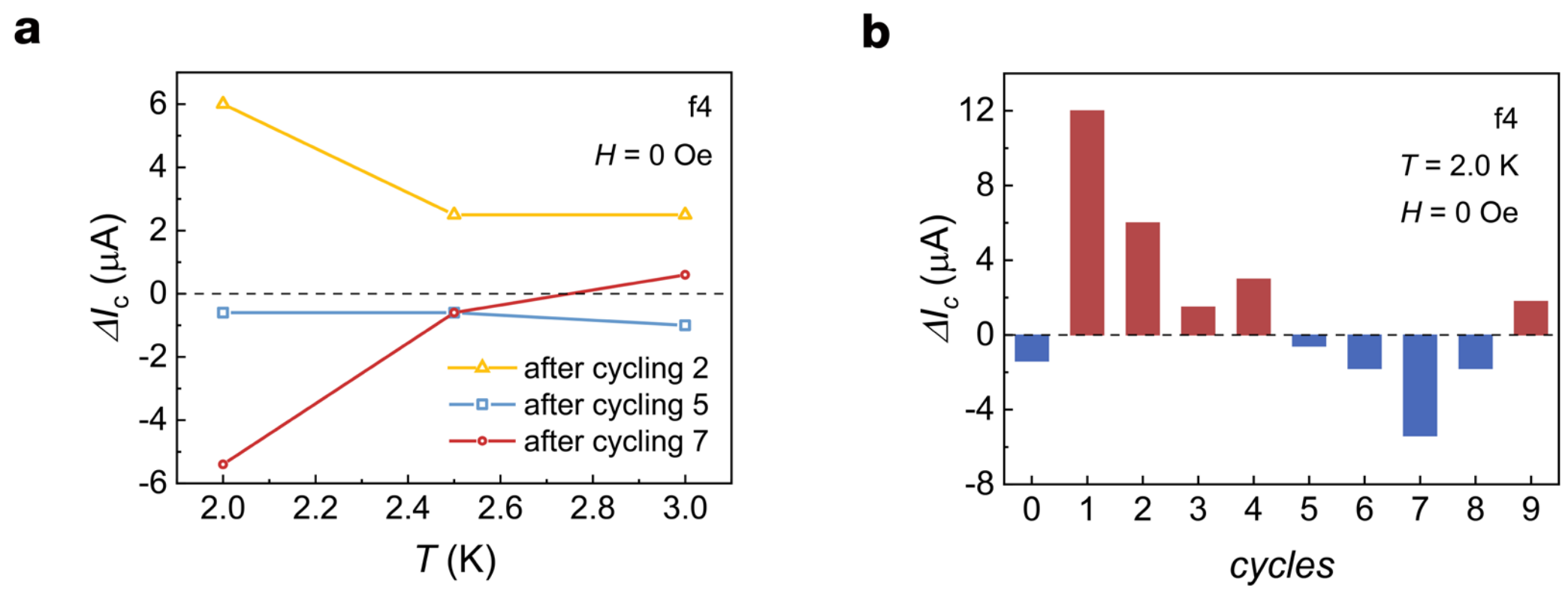

**Supplementary Fig. 9 | The non-reciprocal critical current $\Delta I_c$ in the $CsV_3Sb_5$ flake device f4 under zero-field thermal cycling.** The measurements are performed at 2 K under zero external magnetic field. For each thermal cycle, the temperature is increased to 300 K (well above $T_{CDW}$~70 K) at zero magnetic field and then decreased to 2 K. **a**, The non-reciprocal critical current $\Delta I_c$ as a function of temperature under thermal cycling 2, 5 and 7 in flake device f4. **b**, $\Delta I_c$ as a function of cycle numbers. Stochastic polarity switching is observed under thermal cycles.

**Supplementary Note 1. The correlation between the devices sizes and the stability of the SDE polarity against thermal cycles**

As shown in Supplementary Figs. 9, 10, the flake devices with larger widths (f1 and f4 with dimensions shown in Supplementary Table 2) more frequently exhibit polarity reversal upon thermal cycling compared to smaller flakes (f2, f3). This trend is further supported by observations in our micro-bridge devices, which have the smallest widths (Supplementary Table 3) of all tested samples. In the micro-bridges, the SDE polarity remains constant across the measured temperature range (Fig. 5h in the manuscript). Furthermore, the SDE polarity in the micro-bridges is insensitive to thermal cycling (Supplementary Fig. 11). These results suggest that the geometric confinement in smaller devices may help stabilize the time reversal symmetry breaking (TRSB) domain configuration, reducing the stochastic switching behavior observed in larger flakes.

**Supplementary Table 2 | Geometric parameters and superconducting transition temperatures for the measured $CsV_3Sb_5$ flake devices f1-f4.** Here, *L1* denotes the total sample length, *L2* is the distance between voltage electrodes, *W* is the sample width, and *d* is the sample thickness.

| | L1 (μm) | L2 (μm) | W (μm) | d (nm) | $T_c^{onset}$(K) | $T_c^{zero}$ (K) |
|---|---|---|---|---|---|---|
| f1 | 23.3 | 5.1 | 13.4 | 23.0 | 4.11 | 3.04 |
| f2 | 20.0 | 4.9 | 8.8 | 8.8 | 3.50 | 3.18 |
| f3 | 11.3 | 2.2 | 9.1 | 33.0 | 4.32 | 3.80 |
| f4 | 47.4 | 6.1 | 19.2 | 22.5 | 4.06 | 3.45 |

**Supplementary Table 3 | Geometric parameters and superconducting transition temperatures for the measured $CsV_3Sb_5$ micro-bridge devices s1, s2**. Here, *L* is the bridge length, *W* is the bridge width, and *d* is the sample thickness.

| | L (μm) | W (μm) | d (nm) | $T_c^{onset}$ (K) | $T_c^{zero}$ (K) |
|---|---|---|---|---|---|
| s1 | 2.83 | 0.479 | 27.5 | 4.22 | 3.04 |
| s2 | 2.80 | 0.516 | 20.0 | 3.97 | 2.95 |

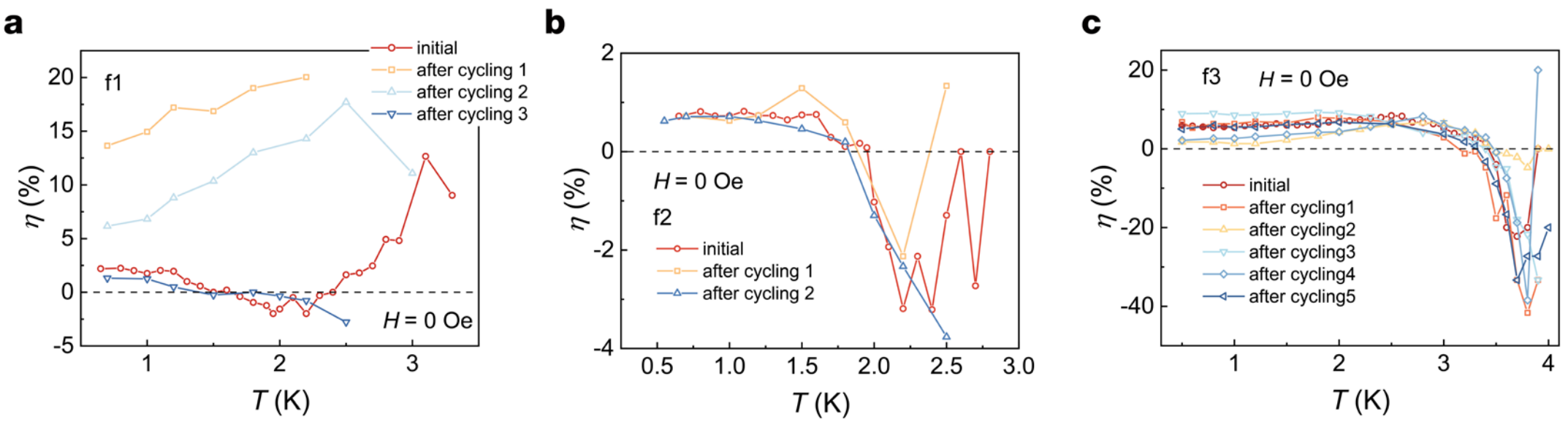

**Supplementary Fig. 10 | The diode efficiency η as a function of temperature before and after zero-field thermal cycles in flake devices f1 (a), f2 (b) and f3 (c).**

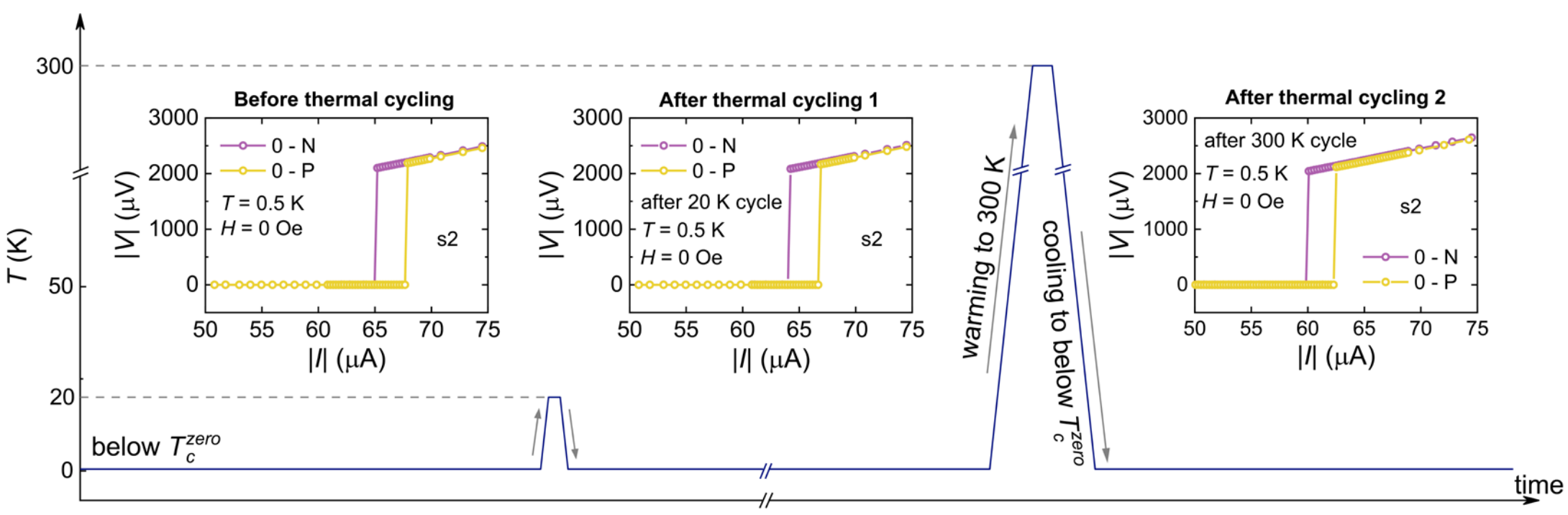

**Supplementary Fig. 11 | The superconducting diode effect (SDE) under zero-field thermal cycles in the $CsV_3Sb_5$ micro-bridge device s2**. The schematic illustration of the thermal cycling process is shown. For thermal cycle 1, the temperature is increased to 20 K (well above the superconducting onset transition temperature but below the charge density wave transition temperature) at zero magnetic field and then decreased to reach the zero-resistance superconducting state. For thermal cycle 2, the temperature is increased to 300 K (well above the charge density wave transition temperature) at

zero magnetic field and then decreased to reach the zero-resistance superconducting state. Insets show the $V$–$I$ curves at 0.5 K before cycling and after cycling 1, 2.

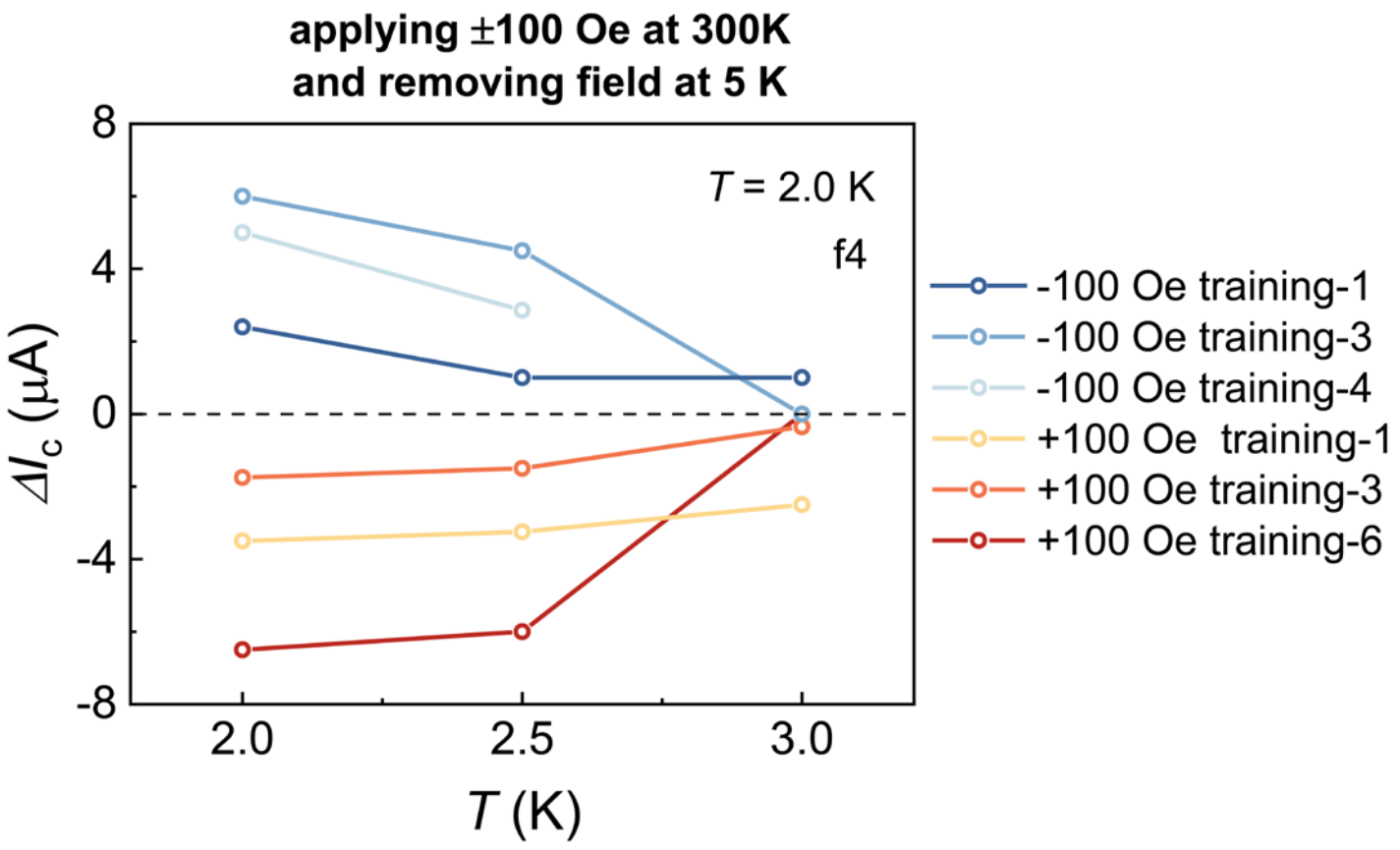

**Supplementary Fig. 12 | Temperature dependence of $\Delta I_c$ for flake device f4 following multiple field training cycles (field applied at 300 K and removed at 5 K).** Different colors represent independent training events. Crucially, for each training cycle, the sign of $\Delta I_c$ remains invariant across the measured temperature range (2.0–3.0 K). This indicates that the specific domain configuration set by the training field is thermally stable and does not undergo polarity reversal as temperature increases, which is in stark contrast to the behavior in the untrained state (Supplementary Fig. 7).

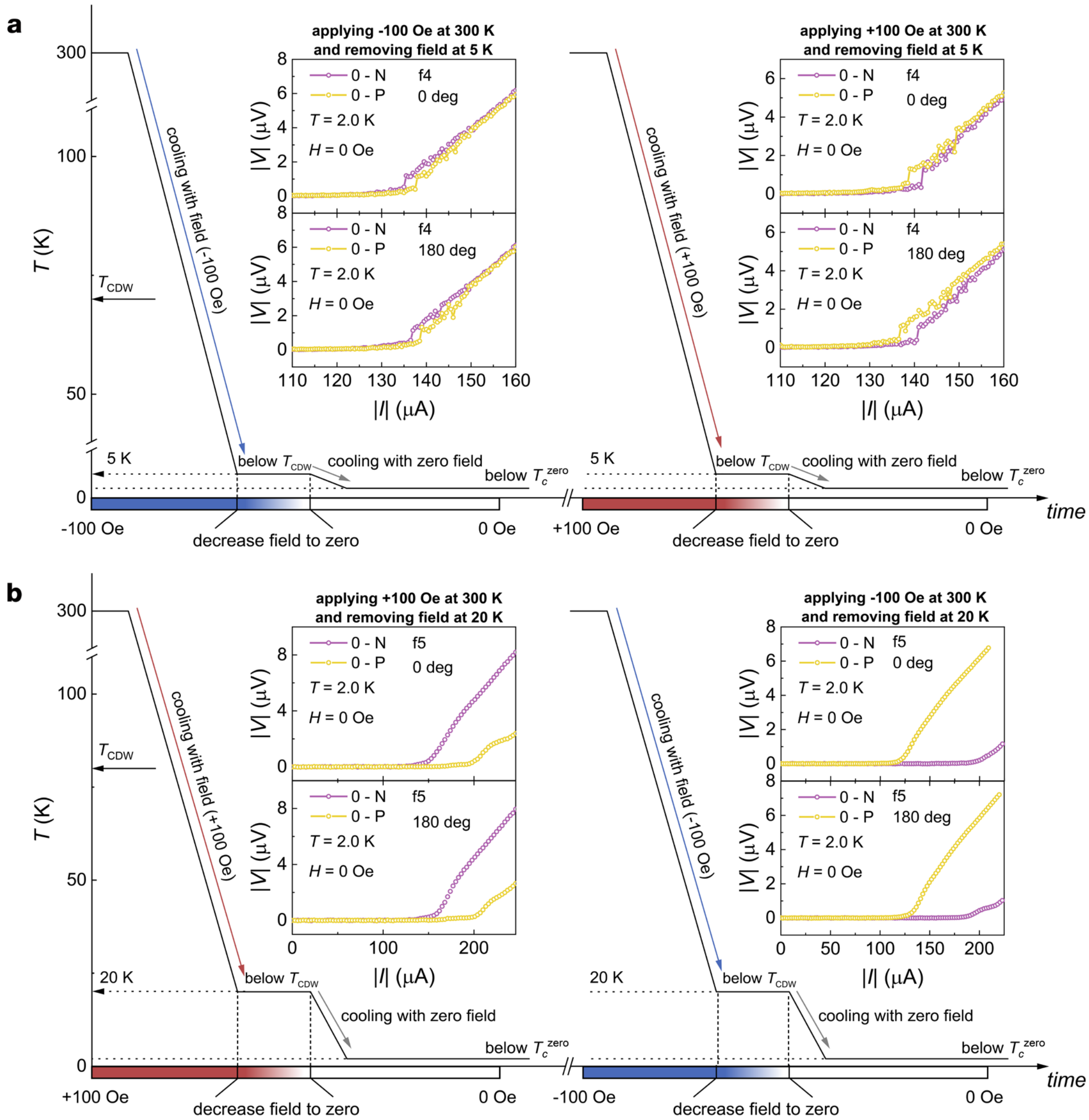

**Supplementary Fig. 13 | SDE observed in flake devices f4 (a) and f5 (b) after field training and sample flipping.** The main panels show the field cooling processes from 300 K under ±100 Oe. The insets display the *V*-*I* curves at 2 K measured at zero field. "0 deg" and "180 deg" correspond to the configurations before and after sample flipping, respectively. The SDE polarity remains unchanged after flipping the devices.

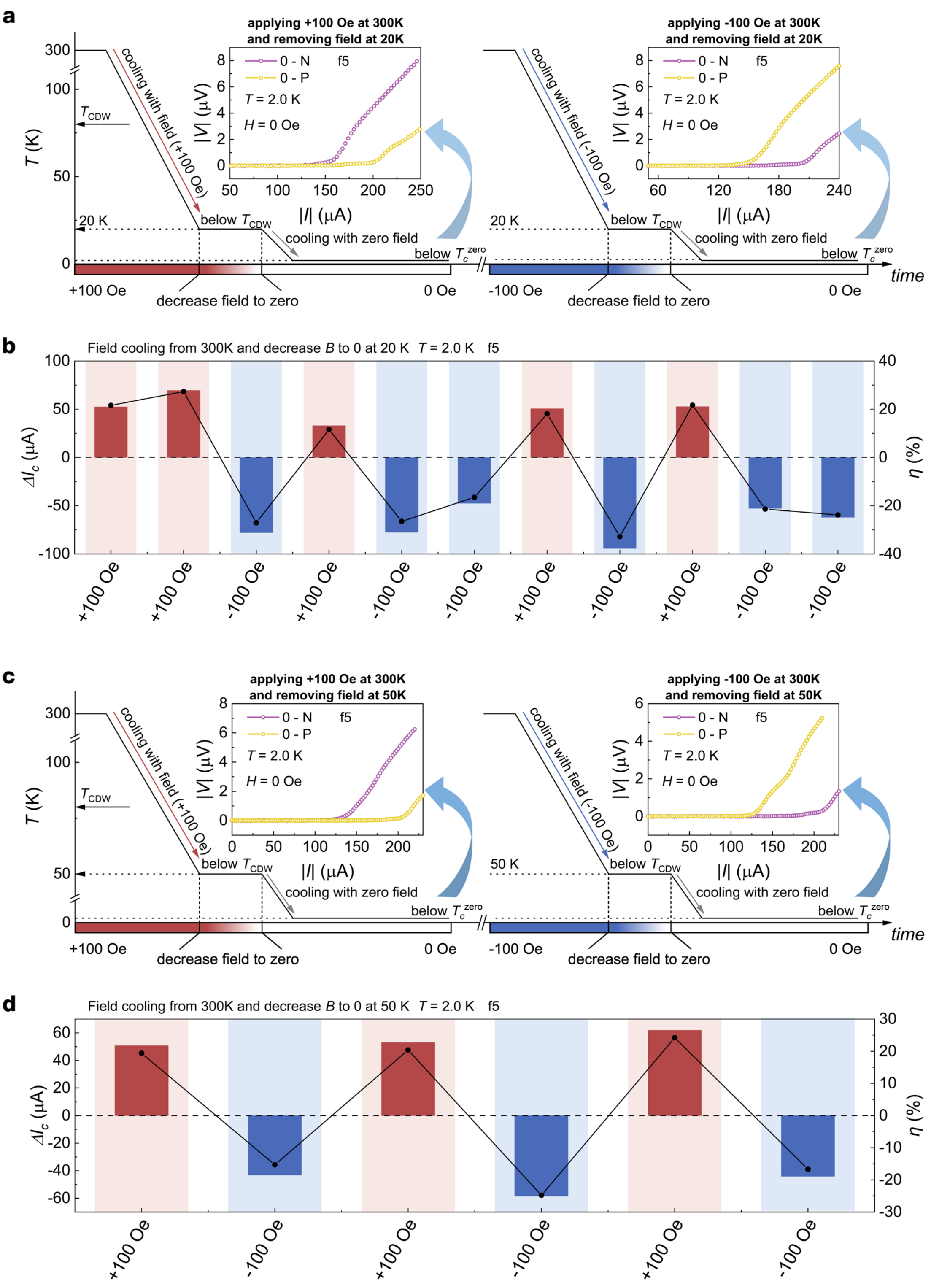

**Supplementary Fig. 14 | Field training on the superconducting diode effect (SDE) in the $CsV_3Sb_5$ flake device f5, with the magnetic field removed at 20 K (a, b) and 50 K (c, d). a, c,** Schematic illustration of the field training process. The left panel shows the procedure under a +100 Oe (indicated in red on the horizontal axis) field training, while the right panel corresponds to −100 Oe (indicated in blue on the horizontal axis) field training. During field training, a magnetic field (+100 Oe or -100 Oe) is first applied at room temperature, followed by cooling the sample down to 20 K

or 50 K under the field. These temperatures are below the charge density wave (CDW) transition temperature ($T_{CDW}$~80 K) of device f5. The magnetic field is then decreased to zero at 20 K or 50 K, and the sample is further cooled to below the zero-resistance temperature, where the *I–V* measurements are performed. Insets show the *V-I* curves measured at 2 K after +100 Oe (left panel) and -100 Oe (right panel) field training. **b, d,** The polarity change at 2.0 K in f5 generated by reversing the field in the normal state. The color and height of the bars represent the sign (dark red for positive, dark blue for negative) and the amplitude of $\Delta I_c$, respectively. The background shading indicates the direction of training field (light red for +100 Oe and light blue for -100 Oe). The black dots correspond to the diode efficiency η (right axis).

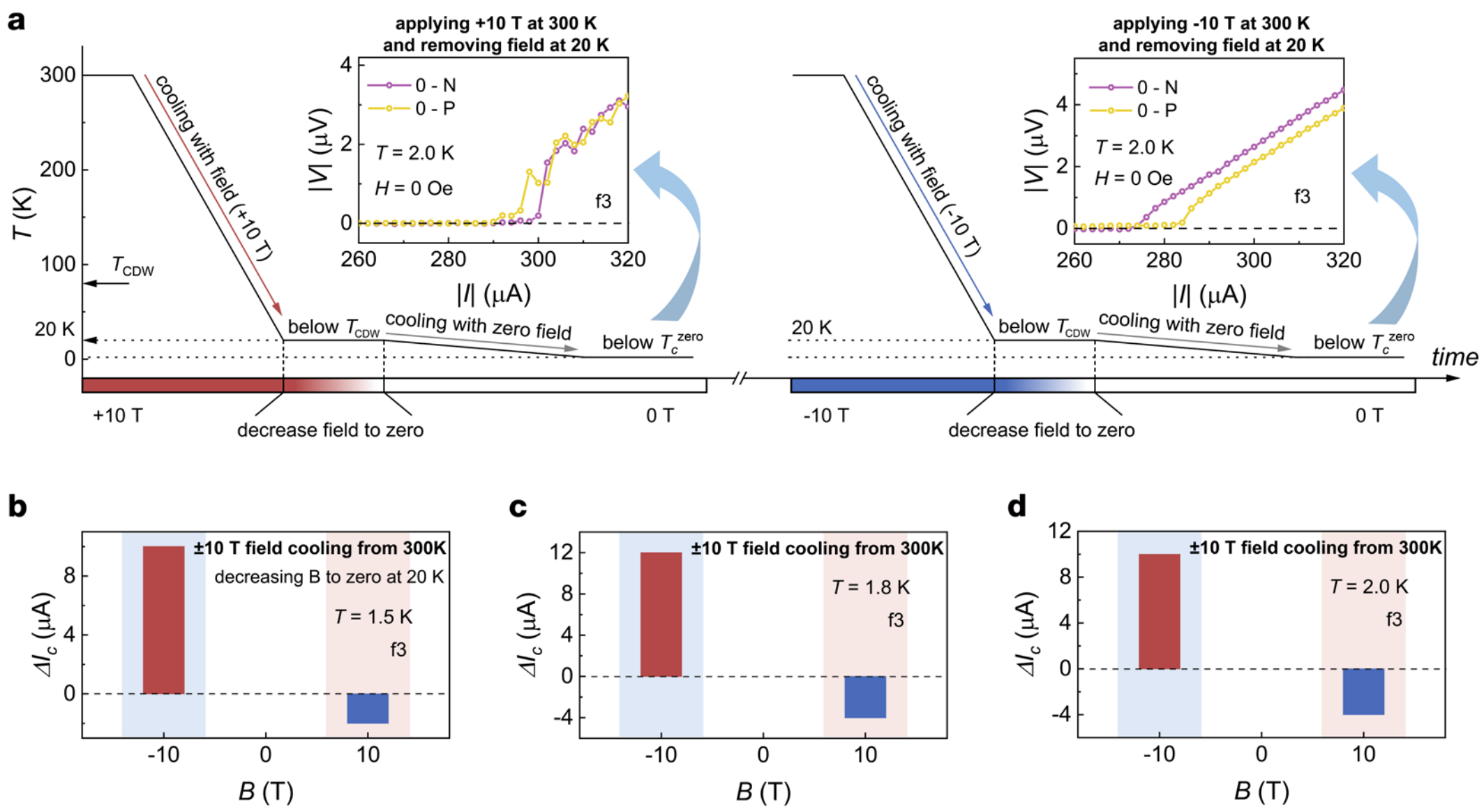

**Supplementary Fig. 15 | Field training on the superconducting diode effect (SDE) in the $CsV_3Sb_5$ flake device f3. a**, Schematic illustration of the field training process. The left panel shows the procedure under a +10 T (indicated in red on the horizontal axis) field training, while the right panel corresponds to −10 T (indicated in blue on the horizontal axis) field training. During field training, a magnetic field (+10 T or -10 T) is first applied at room temperature, followed by cooling the sample down to 20 K under the field. This temperature is below the charge density wave transition temperature ($T_{CDW}$~78 K) of device f3. The magnetic field is then decreased to zero at 20 K, and the sample is further cooled to below the zero-resistance temperature, where the *I–V* measurements are performed. Insets show the *V-I* curves measured at 2 K after +10 T (left panel) and -10 T (right panel) field training. **b-d**, The polarity change at 1.5 K (b), 1.8 K (c) and 2.0 K (d) in f3 generated by reversing the field in the normal state. The color and height of the bars represent the sign (dark red for positive, dark blue for negative) and the amplitude of $\Delta I_c$, respectively. The background shading indicates the direction of training field (light red for +10 T and light blue for -10 T). Note that the field-training-tuned polarity reversal here is not caused by thermal cycling, as the polarities at 1.5 K, 1.8 K and 2.0 K in f3 remain unchanged in thermal cycles (see more details in Supplementary Table 4).

**Supplementary Table 4 | The polarity of the superconducting diode effect (SDE) in flake device f3 after thermal cycling and field training at 1.5 K, 1.8 K, and 2.0 K.** The polarity at 1.5 K, 1.8 K and 2.0 K can only be reversed by field training. The background shading indicates the direction of the training field (light red for +10 T, light blue for -10 T).

| f3 | Initial polarity | After thermal cycling1 | After thermal cycling2 | After thermal cycling3 | After thermal cycling4 | After thermal cycling5 | +10 T field training | -10 T field training |
|---|---|---|---|---|---|---|---|---|
| 1.5 K | + | + | + | + | + | + | - | + |
| 1.8 K | + | + | + | + | + | + | - | + |
| 2.0 K | + | + | + | + | + | + | - | + |

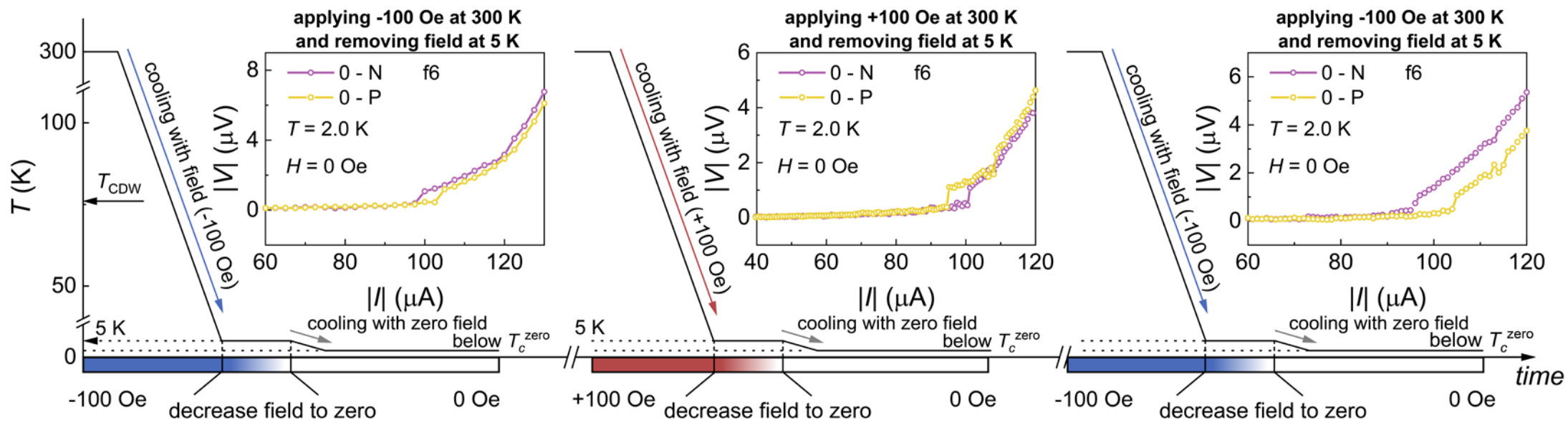

**Supplementary Fig. 16 | Field training on the superconducting diode effect (SDE) in the $CsV_3Sb_5$ flake device f6, with the magnetic field removed at 5 K.** The left panel shows the procedure under a -100 Oe (indicated in blue on the horizontal axis) field training, the middle panel corresponds to +100 Oe (indicated in red on the horizontal axis), and the right panel exhibits another -100 Oe (indicated in blue on the horizontal axis) field training. During field training, a magnetic field (+100 Oe or -100 Oe) is first applied at room temperature, followed by cooling the sample down to 5 K under the field. This temperature is below the charge density wave (CDW) transition temperature ($T_{CDW}$~76 K) of device f6. The magnetic field is then decreased to zero at 5 K, and the sample is further cooled to below the zero-resistance temperature, where the *I*–*V* measurements are performed. Insets show the *V*-*I* curves measured at 2 K after -100 Oe (left panel), +100 Oe (middle panel) and -100 Oe (right panel) field training.

**Supplementary Note 2. Comparison of field training effects initiated above and below the CDW transition temperature**

We compare the SDE results obtained using the standard training procedure (field applied at 300 K, above $T_{CDW}$) with three distinct control experiments performed within the CDW phase (field applied below $T_{CDW}$). The control protocols involve field application and removal at 5 K, field application and removal at 50 K and field application at 50 K and removal at 5 K. As shown in Supplementary Fig. 17, we observe that in all three control scenarios, the polarity consistently switches with the direction of the applied field, mirroring the behavior observed in the 300 K training. This trainability of the SDE polarity by a magnetic field inside the ordered CDW phase serves as further evidence for TRSB in the CDW phase.

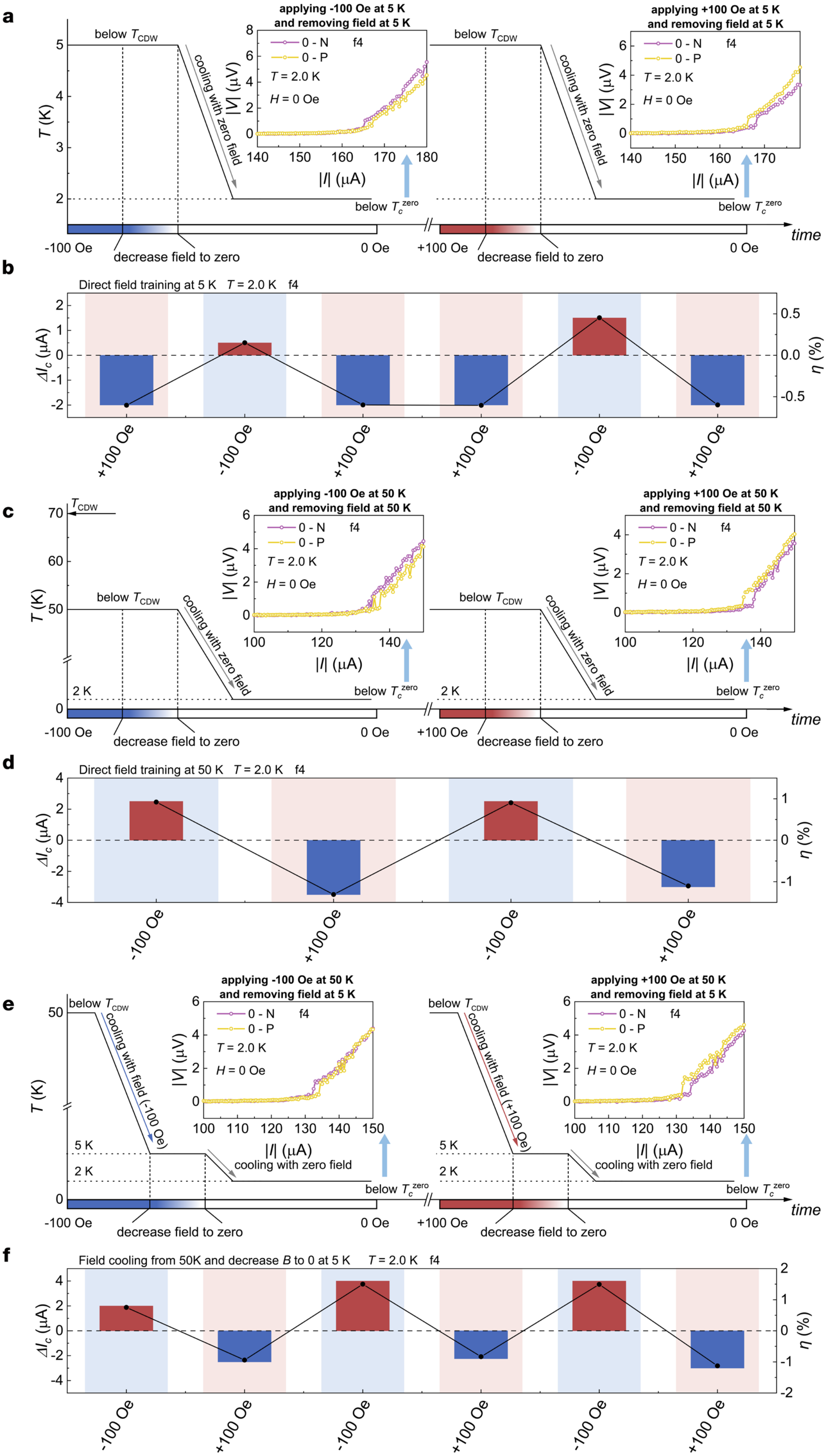
a
below $T_{CDW}$
cooling with zero field
applying -100 Oe at 5 K
and removing field at 5 K
0 - N
0 - P
f4
$T$ = 2.0 K
$H$ = 0 Oe
$|V|$ (μV)
$|I|$ (μA)
below $T_c^{zero}$
applying +100 Oe at 5 K
and removing field at 5 K
$T$ (K)
-100 Oe
decrease field to zero
0 Oe
+100 Oe
time
b
Direct field training at 5 K $T$ = 2.0 K f4
$\Delta I_c$ (μA)
$\eta$ (%)
+100 Oe
-100 Oe
c
$T_{CDW}$
applying -100 Oe at 50 K
and removing field at 50 K
applying +100 Oe at 50 K
and removing field at 50 K
2 K
d
Direct field training at 50 K $T$ = 2.0 K f4
e
cooling with field (-100 Oe)
cooling with field (+100 Oe)
applying -100 Oe at 50 K
and removing field at 5 K
applying +100 Oe at 50 K
and removing field at 5 K
5 K
cooling with zero field
f
Field cooling from 50K and decrease $B$ to 0 at 5 K $T$ = 2.0 K f4

**Supplementary Fig. 17 | Field training on the superconducting diode effect (SDE) in the $CsV_3Sb_5$ flake device f4, with both the magnetic field application and removal performed below $T_{CDW}$.** The experiments consist of three protocols: magnetic field applied and removed at 5 K (**a**, **b**); applied and removed at 50 K (**c**, **d**); and applied at 50 K with removal at 5 K (**e**, **f**). **a**, **c**, **e**, Schematic illustrations of the field training processes. In each schematic, the left panel shows the procedure under a -100 Oe field (blue axis), while the right panel corresponds to a +100 Oe field (red axis). Insets in **a**, **c**, **e** show the *V-I* curves measured at 2 K after -100 Oe (left panel) and +100 Oe (right panel) field training. **b**, **d**, **f**, The SDE polarity observed at 2.0 K corresponding to the training protocols in **a**, **c**, and **e**, respectively. The color and height of the bars represent the sign (dark red for positive, dark blue for negative) and the amplitude of $\Delta I_c$, respectively. The background shading indicates the direction of training field (light red for +100 Oe and light blue for -100 Oe). The black dots correspond to the diode efficiency η (right axis).

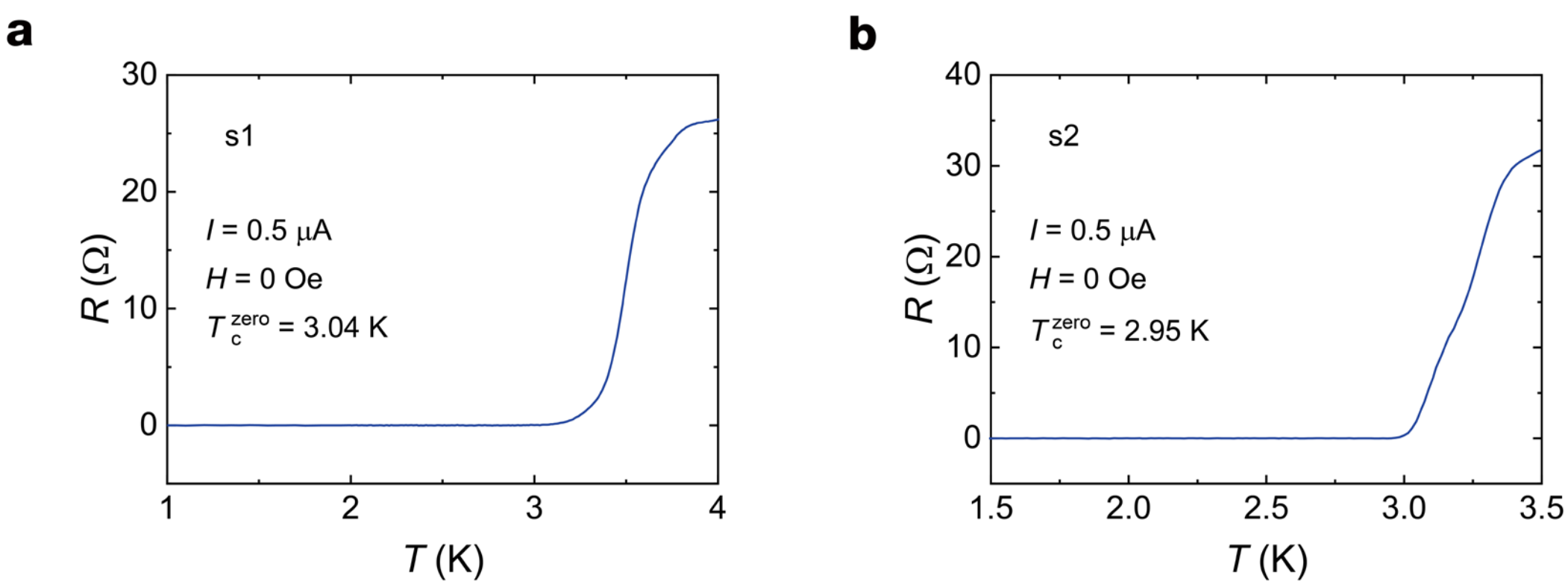

**Supplementary Fig. 18 | Resistance vs. temperature (*R−T*) curves for the micro-bridge devices s1 (a) and s2 (b).**

**Supplementary Note 3. Influence of the geomagnetic field**

We have carefully evaluated the possible influence of the geomagnetic field and confirmed that it does not affect our results. First, our PPMS system has the magnetic shielding, which significantly attenuates the external field, and the remaining vertical component is further effectively eliminated through our active compensation procedure described in the main text. The remaining horizontal component is approximately 0.1 Oe as measured by the Hall sensor, which is parallel to the superconducting flake sample. As discussed below, this residual horizontal field has a negligible effect on the observed SDE.

For the zero-field results prior to field training, the SDE polarity evolves with temperature (Supplementary Fig. 7) and also varies across different thermal cycles (Supplementary Fig. 9). These observations are inconsistent with the static nature of the geomagnetic field since a static field is supposed to induce a SDE with constant polarity. Furthermore, in our sample-flipping experiments (Supplementary Fig. 6), the SDE signal remains unchanged. If the tiny static horizontal background field were responsible for the SDE, reversing the sample orientation relative to this fixed field would be expected to alter the SDE polarity.

For the field-training measurements, a strict one-to-one correspondence between the SDE polarity and the direction of the training magnetic field is observed only when the field is removed below the charge density wave transition temperature. This behavior further confirms that the signal originates from the intrinsic properties of the material rather than from a constant background field.